\documentclass[twocolumn,showpacs,preprintnumbers,amsmath,amssymb]{revtex4}

\usepackage{epsfig}
\usepackage{graphicx}
\usepackage{dcolumn}
\usepackage{bm}

\def\DESepsf(#1 width #2){\epsfxsize=#2 \epsfbox{#1}}
\newcommand{\im}{{\rm Im}}
\newcommand{\out}{{\rm out}}

\begin{document}


\title{Final State Rescattering
and Color-suppressed $\overline B{}^0\to D^{(*)0} h^0$ Decays}

\author{Chun-Khiang Chua}
\author{Wei-Shu Hou}%
\affiliation{%
Physics Department,
National Taiwan University,
Taipei, Taiwan 10764,
Republic of China
}%

\author{Kwei-Chou Yang}
\affiliation{ Physics Department, Chung Yuan Christian University,
Chung-Li, Taiwan 32023, Republic of China
}%

\date{\today}

\begin{abstract}
The color-suppressed
$\overline B{}^0\to D^{(*)0}\pi^0,\,D^{(*)0}\eta,\,D^0\omega$ decay modes
have just been observed for the first time.
The rates are all larger than expected,
hinting at the presence of final state interactions.
Considering $\overline B{}^0\to D^{(*)0}\pi^0$ mode alone, 
an elastic $D^{(*)}\pi \to D^{(*)}\pi$ rescattering phase difference 
$\delta \equiv \delta_{1/2} - \delta_{3/2} \sim 30^\circ$ would suffice, 
but the $\overline B{}^0\to D^{(*)0}\eta,\,D^0\omega$ modes compel one to
extend the elastic formalism to SU(3) symmetry.
We find that a universal $a_2/a_1=0.25$ and two strong phase differences
20$^\circ \sim \theta < \delta < \delta^\prime \sim$ 50$^\circ$ 
can describe both $DP$ and $D^*P$ modes rather well;
the large phase of order $50^\circ$ is needed to account for 
the strength of {\it both} the $D^{(*)0}\pi^0$ and $D^{(*)0}\eta$ modes.
For $DV$ modes, the nonet symmetry reduces the number of physical phases to just one,
giving better predictive power. Two solutions are found.
We predict the rates of the 
$\overline B{}^0\to D^{+}_s K^-$, $D^{*+}_s K^-$, $D^0\rho^0$,
$D^+_s K^{*-}$ and $D^0\phi$ modes, as well as
$\overline B{}^0\to D^{0}\overline K{}^0$, $D^{*0}\overline K{}^0$, 
$D^{0}\overline K{}^{*0}$ modes. 
The formalism may have implications for rates and
$CP$ asymmetries of charmless modes.
\end{abstract}

\maketitle

\section{Introduction}

The Belle Collaboration has recently observed \cite{Abe:2001zi} the
$\overline B{}^0 \to D^0\pi^0$, $D^{*0}\pi^0$, $D^0\eta$ and $D^0\omega$ decay modes,
as well as finding evidence for $\overline B{}^0 \to D^{*0}\eta$ and 
$D^{*0}\omega$.
The decay branching ratios (${\cal B}$) are all at a few times $10^{-4}$ level:
\begin{eqnarray}
{\cal B} (\overline{B}^0 \to D^0 \pi^0) &=&
	(3.1 \pm 0.4 \pm 0.5) \times 10^{-4},
\nonumber \\
{\cal B} (\overline{B}^0 \to D^{*0} \pi^0) &=&
	(2.7 \;^{+0.8}_{-0.7}\;^{+0.5}_{-0.6})\times 10^{-4},
\nonumber \\
{\cal B} (\overline{B}^0 \to D^0 \eta) &=&
	(1.4\;^{+0.5}_{-0.4}\pm 0.3) \times 10^{-4},
\nonumber \\
{\cal B} (\overline{B}^0 \to D^0 \omega) &=&
	(1.8 \pm 0.5 \;^{+0.4}_{-0.3}) \times 10^{-4},
\nonumber \\
{\cal B} (\overline{B}{}^0 \to D^{*0} \eta) &=&
	(2.0\;^{+0.9}_{-0.8} \pm 0.4) \times 10^{-4},
\nonumber \\
{\cal B} (\overline{B}{}^0 \to D^{*0} \omega) &=&
	(3.1\;^{+1.3}_{-1.1}\pm 0.8) \times 10^{-4}.
\nonumber 
\end{eqnarray}
The CLEO Collaboration has also reported \cite{Coan:2001ei}
the observation of $\overline B{}^0 \to D^0\pi^0$, $D^{*0}\pi^0$ modes,
\begin{eqnarray}
{\cal B} (\overline{B}{}^0 \to D^0 \pi^0) &=&
	(2.74\;_{-0.32}^{+0.36} \pm 0.55) \times 10^{-4},
\nonumber \\
{\cal B} (\overline{B}{}^0 \to D^{*0} \pi^0) &=&
	(2.20\;_{-0.52}^{+0.59} \pm 0.79)\times 10^{-4},
\nonumber 
\end{eqnarray}
with rates in agreement with Belle.
These modes are usually called color-suppressed $B$ decays.
In contrast to the much faster ``color-allowed" $\overline B{}^0\to D^+\pi^-$
(${\cal B} \simeq 3\times 10^{-3}$~\cite{Groom:2000in}) decay
where $\pi^-$ is emitted by the charged current (Fig.~1(a)),
there is a color mismatch in forming the $D^0$ meson from
$c$ and $\overline u$ produced by $b\to c\overline ud$ decay (Fig.~1(b)).
We have indicated in Figs. 1(a) and (b)
the effective Wilson coefficients,
$a_1$ and $a_2$ \cite{Neubert:1997uc},
that is responsible for the decay.
For the charged $B^- \to D^0\pi^-$ decay,
both $a_1$ and $a_2$ type of diagrams contribute.

\begin{figure}[b!]
\centerline{\ \DESepsf(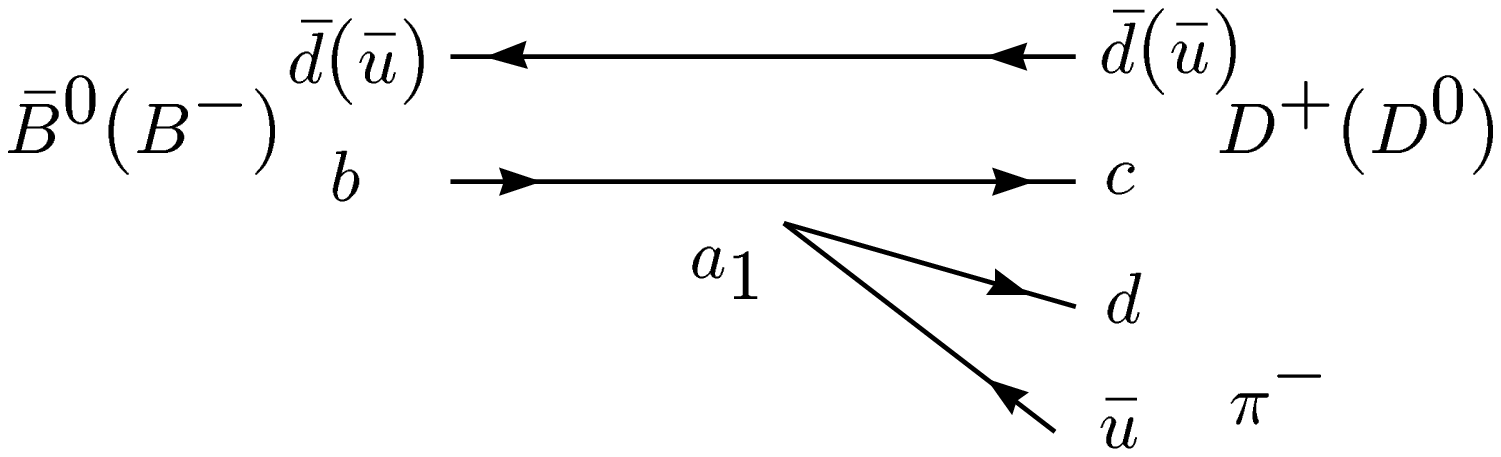 width 7cm)}
\centerline{(a)}
\vskip0.1cm
\centerline{\ \DESepsf(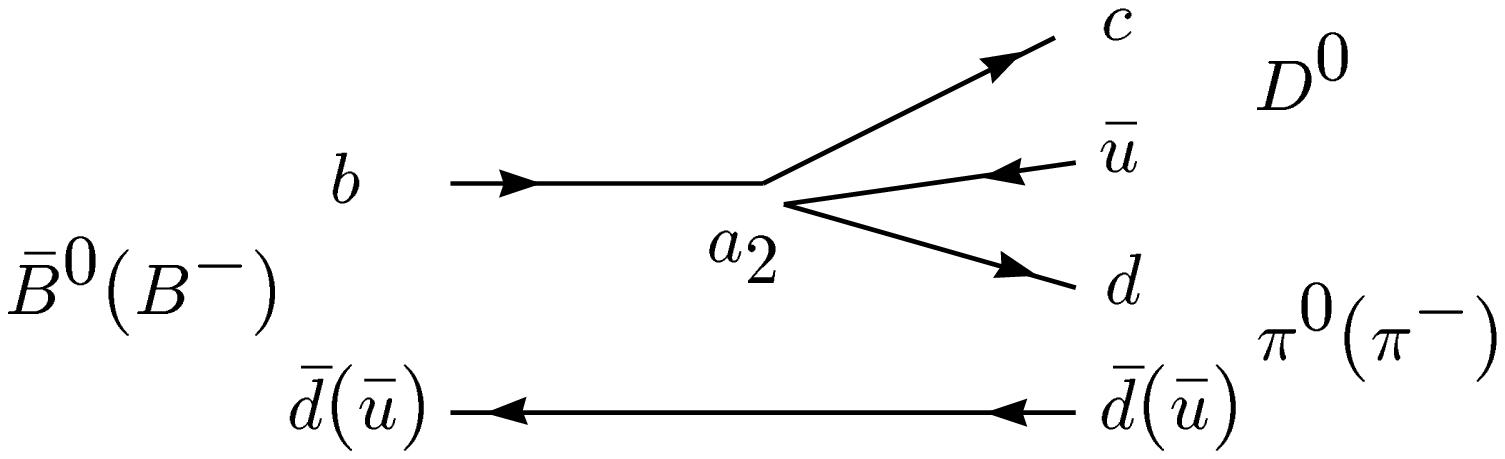 width 7cm)}
\centerline{(b)}
\smallskip
\caption{Color (a) allowed and (b) suppressed $B\to D\pi$ decays.
}
\label{fig:a1a2}
\end{figure}


The factorization of the $\overline B{}^0\to D^+\pi^-$ decay amplitude
has recently been demonstrated to follow from QCD in
the heavy $b$ quark limit \cite{Beneke:2000ry,Bauer:2001cu},
and $a_1$ as computed from QCD factorization is
close to $a_1^{({\rm eff})}\sim 1$--$1.1$ from
``generalized factorization" \cite{Neubert:1997uc}.
On the other hand, it is known that
the effective coefficient $a_2^{({\rm eff})}$
cannot be calculated in QCD factorization \cite{Beneke:2000ry}.
It is remarkable that the $a_2^{({\rm eff})}$ value
as extracted from the celebrated $\overline B{}\to J/\psi \overline K{}^{(*)}$ modes
agrees rather well with the value extracted from $B^- \to D^0\pi^-$,
i.e. consistent with $a_2^{({\rm eff})}\simeq 0.2$--$0.3$ and real.
However, the $D^{(*)0}\pi^0$ rates
as observed by Belle and CLEO are considerably higher than
the generalized factorization estimates \cite{Neubert:1997uc}
using this $a_2^{({\rm eff})}$ value,
suggesting that final state rescatterings (FSI)
such as $D^{(*)+}\pi^- \to D^{(*)0}\pi^0$ could be active.
Alternatively, it could indicate non-universal $a_2$,
that is, $a_2$ for color-suppressed modes are
larger than \cite{Neubert:2001sj} $a_2$ for $J/\psi \overline K{}^{(*)}$ modes,
and are furthermore complex to accommodate $B^- \to D^{(*)0}\pi^-$.
Whether in the form of FSI rescattering or a larger and complex $a_2$,
one in general acquires complexity of decay amplitudes
in the form of $CP$ conserving phases.
This may have implications for direct $CP$ violation asymmetries,
since similar effects may be present in processes such as
charmless $B\to K\pi$ modes.

The experimental data in fact compels one to broaden the horizon.
The Belle results on $\overline B{}^0 \to D^0\eta$ and $D^0\omega$ are
considerably higher than generalized factorization estimates
 \cite{Neubert:1997uc} as well, suggesting that one needs to go beyond
$D\pi \to D\pi$ considerations in the FSI framework.
The alternative of having process dependent $a_2$'s
 \cite{Neubert:2001sj,Cheng:2001sc}
would imply loss of predictive power.
While it is clear that $a_2$ in general will be process dependent,
we do abhor the loss of predictive power.
In this paper we wish to explore how the situation could be remedied.

Since the data is new and still rather incomplete,
our approach will be phenomenological, without aim for rigor or completeness.
Let us start from the isospin decomposition of $\overline B\to D^0\pi^-$,
$D^+\pi^-$ and $D^0\pi^0$ decay amplitudes that is
lucidly outlined in Ref. \cite{Neubert:1997uc},
\begin{eqnarray}
A_{D^0\pi^-} &=&{\sqrt3}A_{3/2},
\nonumber \\
A_{D^+\pi^-} &=&\sqrt{2\over3} A_{1/2}+\sqrt{1\over3}A_{3/2},
\label{eq:isospin}
 \\
A_{D^0\pi^0} &=&\sqrt{1\over3} A_{1/2}-\sqrt{2\over3}A_{3/2},
\nonumber
\end{eqnarray}
where the final state is emphasized.
In the absence of FSI, the smallness of $D^0\pi^0$ rate can be viewed as
due to cancellation between the $A_{1/2}$ and $A_{3/2}$ amplitudes,
which are real under factorization.
But these amplitudes in general become complex under FSI 
and $A_{D^0\pi^0}$ stands to gain strength.
However, the isospin or triangular relation 
$A_{D^+\pi^-}=\sqrt2 A_{D^0\pi^0}+A_{D^0\pi^-}$ always holds.
We note that $A_{D^0\pi^-}$ is purely $A_{3/2}$ and can only rescatter into itself.
It is therefore reasonable to maintain $a_2^{({\rm eff})}$ as 
extracted traditionally, i.e. the same as from $B\to J/\psi K^{(*)}$,
since it is too good to be just a coincidence.

The formalism of Eq.~(\ref{eq:isospin}) can 
generate $\overline B{}^0 \to D^{(*)0}\pi^0$ by {\it elastic}
rescattering from the color-allowed $\overline B{}^0 \to D^{(*)+}\pi^-$
mode \cite{Neubert:1997uc}. 
The problem is the strength of the $\overline B{}^0 \to D^{(*)0}\eta$
and $\overline B{}^0 \to D^{(*)0}\omega$ modes observed by Belle.
The $\eta$ and $\omega$ are isosinglets, and the $\eta$ mass
is quite different from pions.
Thus, to generate $D^{(*)0}\eta$ and $D^{(*)0}\omega$
final states from $D^{(*)+}\pi^-$ and $D^{(*)+}\rho^-$,
strictly speaking, involves {\it inelastic} rescattering,
although only in the isospin 1/2 channel.
However, it may be reasonable to extend from isospin (an excellent symmetry of
the strong interaction) to SU(3), expecting that the latter becomes a
good symmetry at the $m_B$ rescattering scale.

Let us qualify the last statement further.
SU(3) flavor symmetry is a symmetry of the strong interaction as QCD is flavor blind,
except that the flavor symmetry is broken by quark masses.
Thus, in terms of SU(3) multiplets,
masses vary within the multiplet according to SU(3) breaking,
and meson production differs in strength as reflected in,
for example, the decay constants and transition form factors.
But if we enlarge the isospin doublet
$D^{(*)} = (D^{(*)+},\ D^{(*)0})$ and
triplet $\pi = (\pi^+,\ \pi^0,\ \pi^-)$ to
SU(3) triplet D$^{(*)} = (D^{(*)+},\ D^{(*)0},\ D^{(*)+}_s)$
and the meson octet $\Pi$ (likewise from $\rho$ to vector V),
we note that rescattering occurs at $m_B \gg m_q$ scale,
hence the D$\Pi \to$ D$\Pi$ strong rescattering amplitude
should respect SU(3) symmetry to good degree.
SU(3) breaking effects are taken into account in
the initial meson formations from $B$ meson weak decay,
which is done in the (QCD) factorization framework with
$a_1^{({\rm eff})}$ and $a_2^{({\rm eff})}$ as
effective {\it short} distance Wilson coefficients.

Thus, our aim is to extend the usual {\it elastic} $D\pi \to D\pi$ rescattering
to {\it quasi}-elastic D$\Pi \to$ D$\Pi$ rescattering in {\it final state},
so $D^{(*)0}\eta$ and $D^{(*)0}\omega$ 
modes are naturally incorporated.
Our framework is rather close in spirit to
the original isospin analysis of $\overline B{}\to D\pi$,
and is as close to elastic as one can get.
The most general formalism involving inelastic rescatterings 
from all possible hadronic final states cannot be tackled.
In general it involves large cancellations and, 
statistically speaking, small phases \cite{Suzuki:1999uc}.
Hopefully, and in a sense true by duality,
the inelastic effects are contained already in 
$a_1^{({\rm eff})}$ and $a_2^{({\rm eff})}$.

In Sec. II we introduce the general framework of a $T$-matrix,
discuss its link to the optical theorem, and also fix the phase convention.
The formalism is applied to the $\overline B{}\to D\pi$ modes in Sec. III.
Three types of rescattering amplitudes are identified:
a diagonal ``Pomeron"-like piece, and two ``Regge"-like pieces
denoted as ``charge-exchange" and ``annihilation".
We relate these to the usual isospin 1/2 and 3/2 rescattering phases,
and show that only $\delta\equiv \delta_{1/2} - \delta_{3/2}$ matters, 
as expected.

In Sec. IV we extend from SU(2) to SU(3) multiplets in the final state.
For $DP$ and $D^*P$ modes,
where $P$ stands for the pseudoscalar octet, the extension is straightforward.
The question of whether to include the flavor singlet $\eta_1$
in the final state is bypassed by noting
a)~the absence of data, which would remain the case for a while
  unless $\overline B{}^0\to D^0\eta^\prime$ turns out to be
  much larger than $\overline B{}^0\to D^0\eta$, and
b) U$_A(1)$ anomaly that singles out the $\eta_1$ field,
  and perhaps as a consequence,
c) relatively small singlet--octet or $\eta$--$\eta^\prime$ mixing,
  allowing us to identify $\eta_8 \simeq \eta$ for convenience.
We thus ignore $\eta_1$ completely in this work.
We find the previous picture of
three types of rescattering parameters still hold,
but one now has two $\delta$-like phase differences.
The extension to $DV$ final states is treated differently.
By noting that the vector mesons satisfy U$(3)$ rather than SU(3) symmetry,
we use a nonet V field rather than an octet one.
We refrain from discussing $D^*V$ modes since data is scarce,
and since two helicity (or partial wave) amplitudes are involved.

We carry out a numerical study in Sec. V.
For the $DP$ modes, we find two sets of solutions for
the two FSI phase differences,
which are of order $20^\circ$ and $50^\circ$.
One solution is similar to the $D^*P$ case,
and has a very tiny $\overline B{}^0 \to D_s^+ K^-$ decay
due to the smallness of ``annihilation" rescattering.
The other solution gives
${\mathcal B}(\overline B{}^0 \to D_s^+ K^-)\sim5\times10^{-4}$,
which is ruled out by experiment.
For $DV$ modes,
we have two solutions:
one does not have annihilation contribution hence
${\mathcal B}(\overline B{}^0 \to D^0\rho^0)\sim
 {\mathcal B}(\overline B{}^0 \to D^0\omega)$,
while the other does not have exchange contribution and
${\mathcal B}(\overline B{}^0 \to D^0\rho^0) >
{\mathcal B}(\overline B{}^0 \to D^0\omega^0)$.

In Sec. VI we compare our ``$a_2^{({\rm eff})}$ plus FSI rescattering"
approach with the viewpoint of ``process dependent $a_2$",
and discuss possible future applications.
The conclusion is then offered, followed by
Appendices that give same results from a SU(3) decomposition approach
and a geometric (triangular) representation
of our rescattering picture.

\section{Final State Rescattering Framework}

Let $H_{\rm W}$ denote the weak decay Hamiltonian.
We assume absence of weak phases (or they are factored out),
hence, from time reversal invariance of $H_{\rm W}(=U_T H^*_{\rm W} U^\dagger_T$),
one has,
\begin{equation}
\langle i;\out| H_{\rm W}|B\rangle^*
=\sum_j S^*_{ji} \langle j;\out|H_{\rm W}|B\rangle,
\label{eq:timerev}
\end{equation}
where $S_{ij}\equiv\langle i;\out|\,j;{\rm in}\rangle$ is
the strong interaction S-matrix element, and we have used
$U_T|{\rm out\,(in)}\rangle^*=|{\rm in\,(out)}\rangle$
which also fixes the phase convention.
Eq. (\ref{eq:timerev}) can be solved by
(see, for example \cite{Suzuki:1999uc})
\begin{equation}
\langle i; \out|H_{\rm W}|B\rangle=\sum_l S^{1/2}_{il} A^0_{l},
\label{eq:solution}
\end{equation}
where $A^0_{l}$ is a real amplitude.
To show that this is indeed a solution of Eq. (\ref{eq:timerev}),
one needs to use $S_{ij}=S_{ji}$, which follows from time reversal invariance of
the strong interactions and the phase convention we have adopted.
The weak decay amplitude picks up strong scattering phases \cite{Watson:1952ji}.
Also note that since $S^{1/2}$ is unitary, we must have
\begin{equation}
\sum_i |\langle i; \out|H_{\rm W}|B\rangle|^2=\sum_l |A^0_{l}|^2.
\label{eq:unitary}
\end{equation}

Eq. (\ref{eq:timerev}) implies an identity related to
the optical theorem.
Noting that $S=1+i T$, we find
\begin{equation}
2\,\im \langle i;\out|H_{\rm W}|B\rangle
=\sum_j T^*_{ji} \langle j;\out|H_{\rm W}|B\rangle.
\label{eq:ImA}
\end{equation}
Thus, for B decay to a two body final state with momentum ($p_1,p_2$),
one has the relation
\begin{eqnarray}
&&2\,{\rm Im}\,M(p_B\to p_1 p_2)
\nonumber \\
&=& \sum_j\left(\prod_{k=1}^j \int {d^3q_k\over (2\pi)^3 2 E_k}\right)
             (2\pi)^4 \delta^4 \left(p_1+p_2-\sum^j_{k=1} q_k\right)
\nonumber \\
&&\qquad\qquad \times M(p_B\to \{q_k\}) M^*(p_1 p_2\to \{q_k\}),
\label{eq:optical}
\end{eqnarray}
which relates the imaginary part of the two body decay amplitude to
the sum over {\it all possible} $B$ decay final states $\{q_k\}$,
followed by $\{q_k\}\to p_1p_2$ rescattering.
This equation is consistent with the optical theorem
to all orders of the strong interactions but
only to first order of the weak decay vertex,
as we illustrate in Fig. 2.

\begin{figure}[tb]
%
\centerline{\DESepsf(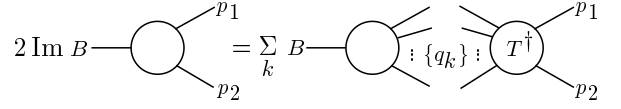 width 8cm)}
\smallskip
\caption{Illustration of optical theorem, Eq. (\ref{eq:optical}).
}
\label{fig:optical1}
\end{figure}


%

Before we turn to $B\to D\pi$ decay applications,
let us give the isospin structure of the related amplitudes.
The responsible effective weak Hamiltonian is given by
\begin{eqnarray}
H_{\rm W}&=&{G_F\over\sqrt2} V_{cb} V^*_{ud}
          \big[c_1\,\overline d \gamma^\mu(1-\gamma_5)u\,
          \overline c \gamma_\mu(1-\gamma_5)b
\nonumber\\
&&+ c_2\,\overline c \gamma^\mu(1-\gamma_5)u\,
          \overline d \gamma_\mu(1-\gamma_5)b\big],
\end{eqnarray}
where $V_{cb} V^*_{ud}$ can be treated as real for our purpose. 
It is clear that $H_{\rm W}$ transforms like a $I=1, I_z=1$ vector
under isospin \cite{Neubert:1997uc}.
It is useful to state explicitly the phase convention and
the isospin structure of these mesons:
\begin{eqnarray}
|\overline B{}^0\rangle&=&|b\overline d\rangle
\sim\overline b{} d|0\rangle\sim\bigg|{1\over2},-{1\over2}\bigg\rangle,
\nonumber \\
|B^-\rangle&=&|b\overline u\rangle
\sim\overline b{} u|0\rangle\sim\bigg|{1\over2},+{1\over2}\bigg\rangle,
\nonumber \\
|D^+\rangle&=&|c\overline d\rangle
\sim \overline c d|0\rangle
\sim\bigg|{1\over2},-{1\over2}\bigg\rangle,
\nonumber \\
|D^0\rangle&=&|c\overline u\rangle\sim \overline c u|0\rangle
\sim\bigg|{1\over2},+{1\over2}\bigg\rangle,
\nonumber \\
|\pi^-\rangle&=&|d\overline u\rangle
\sim \overline d u|0\rangle
\sim |1,+1\rangle,
\nonumber\\
|\pi^0\rangle&=&\bigg|{u\overline u-d\overline d\over \sqrt2}\bigg\rangle
\sim{u\overline u-d\overline d\over \sqrt2}|0\rangle
\sim-|1,\,\,\,0\rangle.
\label{eq:isospinstructure}
\end{eqnarray}
Note that the isospin structure is defined
according to the fields that create the states,
such that they conform with the definition of
the isospin structure of $H_{\rm W}$.  
One could alternatively define isospin quantum numbers according to states
and modify that of $H_{\rm W}$ accordingly.

By isospin decomposition, we have \cite{Neubert:1997uc} (cf. Eq. (\ref{eq:isospin})),
\begin{eqnarray}
A_{D^+\pi^-}&\equiv&\left\langle D^+\pi^-;\out|H_{\rm W}|\overline B{}^0\right\rangle
\nonumber \\
&=&\sqrt{2\over3} A_{1/2}+\sqrt{1\over3}A_{3/2},
\label{eq:isospina}
 \\ 
A_{D^0\pi^0}&\equiv&\left\langle D^0\pi^0;\out|H_{\rm W}|\overline B{}^0\right\rangle
\nonumber \\
&=&\sqrt{1\over3} A_{1/2}-\sqrt{2\over3}A_{3/2},
\label{eq:isospinb}
 \\ 
A_{D^0\pi^-}&\equiv&\left\langle D^0\pi^-;\out|H_{\rm W}|B^-\right\rangle
\nonumber \\
&=&{\sqrt3}A_{3/2},
\label{eq:isospinc}
\end{eqnarray}
where
\begin{eqnarray}
A_{1/2}&=&\left\langle (D\pi)_{1/2};\out|H_{\rm W}|\overline B{}^0\right\rangle,
\nonumber \\
A_{3/2}&=&\left\langle (D\pi)_{3/2};\out|H_{\rm W}|\overline B{}^0\right\rangle
\nonumber \\
       &=&{1\over\sqrt3}\langle (D\pi)_{3/2};\out|H_{\rm W}|B^-\rangle.
\end{eqnarray}
The last step follows from the Wigner-Eckart theorem.
Eqs.~(\ref{eq:isospina})--(\ref{eq:isospinc})
imply the
triangular isospin relation
\begin{equation}
A_{D^+\pi^-}=\sqrt2 A_{D^0\pi^0}+A_{D^0\pi^-}.
\label{eq:triangle}
\end{equation}
We note that the sign of $A_{D^0\pi^0}$ in Eqs. (\ref{eq:isospinb})
and (\ref{eq:triangle}) is different from Ref. \cite{Neubert:1997uc},
but is consistent with Ref.~\cite{Cheng:2001sc}.
It is easy to see from above equations 
that $|A_{1/2}|^2+|A_{3/2}|^2=|A_{D^+\pi^-}|^2+|A_{D^0\pi^0}|^2$.

The isospin relations Eqs. (\ref{eq:isospina})--(\ref{eq:triangle})
are valid whether one has (in)elastic FSI or not.
For example, assuming factorization hence ignoring FSI, one has
\begin{eqnarray}
A_{D^0\pi^0}&\to&A^f_{D^0\pi^0}={1\over \sqrt2}(-{\it C}+{\it E}),
\nonumber \\
A_{D^0\pi^-}&\to&A^f_{D^0\pi^-}={\it T}+{\it C},
\nonumber \\
A_{D^+\pi^-}&\to&A^f_{D^+\pi^-}={\it T}+{\it E},
\label{eq:Af}
\end{eqnarray}
where ${\it T},\,{\it C},\,{\it E}$ are the color-allowed external
$W$-emission, color-suppressed internal $W$-emission and
$W$-exchange amplitudes, which we shall discuss later. 
These amplitudes clearly satisfy Eq. (\ref{eq:triangle}).

The general validity of Eqs.~(\ref{eq:isospina})--(\ref{eq:triangle})
in fact allows one to extract $A_{1/2}$ and $A_{3/2}$ directly from 
the measured $D^{(*)}\pi$ rates without any further recourse to theory. 
Using the Belle and CLEO average of 
${\cal B} (\overline{B}^0 \to D^0 \pi^0) = (2.9 \pm 0.5) \times 10^{-4}$ and
${\cal B} (\overline{B}^0 \to D^{*0} \pi^0) = (2.5 \pm 0.7) \times 10^{-4}$, 
with other rates and
$\tau(B^+)/\tau(B^0)=1.073\pm0.027$ taken from PDG \cite{Groom:2000in},
we find
\begin{eqnarray}
|A_{1/2}|^{\rm expt.}\over\sqrt2\,|A_{3/2}|^{\rm expt.}
&=&0.71\pm0.11\ (0.75\pm0.08),
\nonumber \\
|\delta|^{\rm expt.}
&=&29^\circ\pm6^\circ\ (30^\circ\pm7^\circ),
\label{eq:isospinexpt}
\end{eqnarray}
for $D^{(*)}\pi$ modes, where $\delta\equiv \delta_{1/2}-\delta_{3/2}$
is the phase difference between $A_{1/2}$ and $A_{3/2}$.
This strongly suggests the presence of FSI 
  \cite{Neubert:2001sj,Cheng:2001sc,Xing:2001nj}.

As we turn on FSI, although the isospin relations still hold,
we would clearly lose control if the full structure shown in
Eq. (\ref{eq:optical}) is employed.
Even if all possible $B$ decay rates can be measured,
it would be impossible to know the phases of each amplitude.
Furthermore, we know very little about the strong rescattering amplitudes.
However, the subset of
two body final states that may be reached via
{\it elastic} rescatterings stand out compared to
inelastic channels.
It has been shown from duality arguments \cite{Gerard:1991ni}
as well as a statistical approach \cite{Suzuki:1999uc} that
inelastic FSI amplitudes tend to cancel each other
and lead to small FSI phases.
We shall therefore separate $\{q_k\}$ into
two body elastic channels plus the rest.
We first explore the familiar \cite{Neubert:1997uc}
$\overline B{}\to D^+\pi^-$, $D^0\pi^-$ and $D^0\pi^0$ case,
then try to stretch the scope of elasticity.

\begin{figure}[b]
%
\centerline{\DESepsf(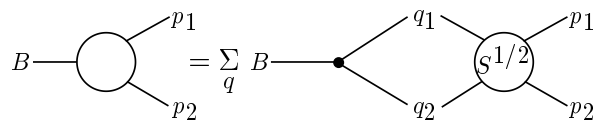 width 8cm)}
\smallskip
\caption{Illustration of Eq. (\ref{eq:FSI}):
two body rescattering.
}
\label{fig:optical2}
\end{figure}

\section{Elastic FSI in the $D\pi$ system}

Let us consider elastic final state rescattering in $\overline B\to D\pi$ modes.
By using Eq. (\ref{eq:solution}), with $A^0_l$ taken as
the factorization amplitudes of Eq. (\ref{eq:Af}), one has
\begin{equation}
\left(
\begin{array}{l}
A_{D^0\pi^-}\\
A_{D^+\pi^-}\\
A_{D^0\pi^0}
\end{array}
\right)
={\cal S}^{1/2}\,
\left(
\begin{array}{l}
A^f_{D^0\pi^-}\\
A^f_{D^+\pi^-}\\
A^f_{D^0\pi^0}
\end{array}
\right).
\label{eq:FSI}
\end{equation}
A major assumption is involved here:
we assume that one can separate hard and soft effects.
The factorization amplitudes sum over ``hard" contributions,
including attempts at incorporating the effects of
the largely intractable full set (generalized to $n$ body) of 
inelastic amplitudes, illustrated for two body final state in Fig.~2.
The ${\cal S}^{1/2}$ matrix describes the
nonperturbative FSI from the factorized $D\pi$ ``source" amplitudes.
We illustrate this in Fig. 3, where the ``hard" part 
is shrunk to a point, and we focus on two body elastic FSI.

It is instructive to show how ${\cal S}^{1/2}$ can be obtained.
In the usual approach, one notes that
with only elastic $D\pi\to D\pi$ rescatterings,
the S-matrix is diagonal in the isospin basis (isospin invariance),
that is
\begin{eqnarray}
{\cal S}_{\rm diag} &=&U\,{\cal S}\,U^{\rm T}
 = {\rm diag} (S_{{3\over2},{3\over2}},\;S_{{3\over2},{3\over2}},
                                       \;S_{{1\over2},{1\over2}}),
\\
U&=&
\left(
\begin{array}{crr}
1 &0 &0\\
0 & \sqrt{1\over3} &-\sqrt{2\over3}\\
0 & \sqrt{2\over3} &\sqrt{1\over3}
\end{array}
\right).
\label{eq:U}
\end{eqnarray}
Unitarity then implies that the diagonal elements of ${\cal S}$
can only be pure phases, or
\begin{equation}
{\cal S}_{{1\over2},{1\over2}}=e^{2i\delta_{1/2}},\quad
{\cal S}_{{3\over2},{3\over2}}=e^{2i\delta_{3/2}}.
\label{eq:Sisospin}
\end{equation}
${\cal S}^{1/2}$ is likewise diagonal,
i.e. $({\cal S}^{1/2})_{{1/2},{1/2}}=e^{i\delta_{1/2}}$~and
$({\cal S}^{1/2})_{{3/2},{3/2}}=e^{i\delta_{3/2}}$.
Elastic FSI is equivalent to $A_I=e^{i\delta_I}|A^f_I|$ 
with $|A^f_I|$ taken from factorization approach \cite{Neubert:1997uc}.
The isospin relation of Eq.~(\ref{eq:triangle}) is clearly satisfied.

For later use we express ${\cal S}^{1/2}$ 
in the basis of Eq. (\ref{eq:FSI}),
\begin{eqnarray}
&&{\cal S}^{1/2}=
 e^{i\delta_{3/2}} \left(
\begin{array}{ccc}
1 & 0  & 0 \\
0 & {1\over 3}(1+{2}e^{i\delta})    &-{\sqrt2\over 3} (1-e^{i\delta}) \\
0 &-{\sqrt2\over 3} (1-e^{i\delta}) & {1\over 3}(2+e^{i\delta})
\end{array}
\right),
\nonumber \\
\label{eq:S12}
\end{eqnarray}
with overall phase $e^{i\delta_{3/2}}$ (i.e. of $A_{D^0\pi^-}$) factored out,
and $\delta$ is the phase difference between isospin 1/2 and 3/2 amplitudes, 
which is physical.

An alternative way to obtain ${\cal S}^{1/2}$ is through the optical theorem,
i.e. Eqs.~(\ref{eq:ImA}) and (\ref{eq:optical}).
This approach is less familiar and more awkward than the previous one.
However, it has the advantage of being readily generalizable to
encompass other modes such as $\overline B{}^0\to D^0\eta, D^+_s K^-$,
which are now of interest
and also provide physical interpretation.
To use the optical theorem, we need to study the strong $S$ matrix,
or equivalently the $T$ matrix. The $T$-operator corresponding to
the matrix element $M(p_1p_2\to q_1q_2)$ for $D\pi\to D\pi$
scattering in Eqs.~(\ref{eq:ImA}) and (\ref{eq:optical}) can be
written as
\begin{eqnarray}
T&=&
\prod_{i,j=1,2} \int {d^3p_i\over (2\pi)^3 2 E_i} {d^3q_j\over (2\pi)^3 2 E_j}
\nonumber\\
&&\times (2\pi)^4 \delta^4(p_1+p_2-q_1-q_2)
\; {\cal D}^\dagger {\cal M} {\cal D}^\prime,
\label{eq:operator}
\end{eqnarray}
where
\begin{eqnarray}
{\cal D}^\dagger &=& \big(a_{D^0}({p_1})\,\ a_{D^+}({p_1})\big),
\,\,\,
{\cal D}^{\prime \rm T}= \big(a^\dagger_{D^0}({q_1})\,\ a^\dagger_{D^+}({q_1})\big),
\nonumber\\
{\cal M}&=&M_0(q_1q_2;p_1 p_2)\,{\rm
Tr}\big(\overline\Pi\,\Pi\big)\,{\bf 1}
\nonumber\\
&&
+M_a(q_1q_2;p_1p_2)\,\Pi\,\overline\Pi
+M_e(q_1q_2;p_1p_2)\,\overline\Pi\,\Pi,
\nonumber \\
\Pi&=&
\left(
\begin{array}{rr}
{1\over\sqrt2}a_{\pi^0}({p_2}) &a_{\pi^+}({p_2})\\
a_{\pi^-}({p_2})              &-{1\over\sqrt2}a_{\pi^0}({p_2})
\end{array}
\right),
\nonumber \\
\overline \Pi&=&
\left(
\begin{array}{rr}
{1\over\sqrt2}a^\dagger_{\pi^0}({q_2}) &a^\dagger_{\pi^-}({q_2})\\
a^\dagger_{\pi^+}({q_2})              &-{1\over\sqrt2}a^\dagger_{\pi^0}({q_2})
\end{array}
\right),
\label{eq:pipibar}
\end{eqnarray}
and $a_M,\,a^\dagger_M$ are annihilation and creation
operators for the meson $M$, respectively.
Eq.~(\ref{eq:operator}) is the most general isospin invariant
operator for $D(p_1)\pi(p_2)\to D^\prime(q_1)\pi^\prime(q_2)$ scattering.
The $T$-operator is defined such that the familiar relation of 
$T$ matrix (used in Eq. (\ref{eq:ImA})) and 
amplitude $M$ (used in Eq. (\ref{eq:optical})) can be reproduced:
\begin{eqnarray}
\langle q_1q_2| T
|p_1p_2\rangle&=&(2\pi)^4\delta^4(q_1+q_2-p_1-p_2)
\nonumber \\
&& \times M(q_1q_2;p_1p_2).
\end{eqnarray}
%

Eq.~(\ref{eq:operator}) is unfamiliar as it is expressed directly in
creation and annihilation operators.
It becomes more familiar when expressed in fields,
where we separate out creation and annihilation parts.
For example, $\Pi$ and $\overline\Pi$ correspond to
the annihilation and creation parts of
$$
\left(
\begin{array}{rr}
{1\over\sqrt2}\pi^0 &\pi^+\\
\pi^-              &-{1\over\sqrt2}\pi^0
\end{array}
\right),
$$
respectively.
The SU(2) transformations of $\Pi$ and $\overline\Pi$ can be recognized,
since they do not mix creation and annihilation parts.
One can find examples of using creation and annihilation operators
on the studies of $\pi\pi$, $\pi$-nucleon scattering in,
for example, Ref. \cite{georgi}.

It is important to note that the $D(p_1)\pi(p_2)\to
D^\prime(q_1)\pi^\prime(q_2)$ scattering amplitudes $M(D\pi\to
D^\prime\pi^\prime)$ can be decomposed into the independent
amplitudes $M_{0,a,e}(q_1q_2;p_1p_2)$. For example, by using Eqs.
(\ref{eq:operator}) and (\ref{eq:pipibar}),
we have
\begin{eqnarray}
M(D^0\pi^-\to D^0\pi^-) &=& M_0+M_e, \nonumber\\
M(D^+\pi^-\to D^+\pi^-) &=& M_0+M_a,
\nonumber\\
M(D^+\pi^-\to D^0\pi^0) &=& {1\over\sqrt2}[M_a-M_e] 
\nonumber\\
&=& M(D^0\pi^0\to D^+\pi^-),
\nonumber\\
M(D^0\pi^0\to D^0\pi^0) &=& M_0+{1\over2}[M_a+M_e],
\label{eq:M1M2}
\end{eqnarray}
where $M_i$ stand for $M_i(q_1q_2;p_1p_2)$.
We can now make use of Eq.~(\ref{eq:optical}) to obtain $\im\,M$.
For example,
\begin{eqnarray}
&&2\,\im\,A_{D^0\pi^-}=
2\,{\rm Im}\,M(B^-\to D^0\pi^-)
\nonumber \\
&&=\int {d^3q_1\over (2\pi)^3 2 E_1}{d^3q_2\over (2\pi)^3 2 E_2}
                                    (2\pi)^4 \delta^4 (p_1+p_2-q_1-q_2)
\nonumber \\
&&\qquad\times M^*(D^0\pi^-\to D^0\pi^-) \; A_{D^0\pi^-}
\nonumber \\
&&=\int {d^3q_1\over (2\pi)^3 2 E_1}{d^3q_2\over (2\pi)^3 2 E_2}
                                    (2\pi)^4 \delta^4 (p_1+p_2-q_1-q_2)
\nonumber \\
&&\qquad \times\big[M^*_0(q_1q_2;p_1p_2)+M^*_e(q_1q_2;p_1p_2)\big]
		 \; A_{D^0\pi^-}
\nonumber \\
&&=(r^*_0+r^*_e)\,A_{D^0\pi^-},
\label{eq:imArA}
\end{eqnarray}
where
\begin{eqnarray}
&&\!\!\!\!\!\!\!\!\!\!\!\!\!\!\!\!\! r^*_{i}\equiv \int
{d^3q_1\over (2\pi)^3 2 E_1}{d^3q_2\over (2\pi)^3 2 E_2}
\nonumber \\
&&\!\!\!\!\!\times(2\pi)^4\delta^4
(p_1+p_2-q_1-q_2)\,M^*_{i}(q_1q_2;p_1p_2),
\label{eq:r1r2def}
\end{eqnarray}
for $i=0,a,e$ and $p_1+p_2=p_B$. 

Since the $D\pi$ system from $B$ decay is S-wave, 
$A_{D\pi}$ is independent of the final state momentum.
It can hence be factored out from the integration. 
Thus, Eq. (\ref{eq:r1r2def})
projects out the S-wave $D\pi$ rescattering amplitude. 
Similar expressions can be obtained for $\im\,A_{D^+\pi^-,D^0\pi^0}$. 
By comparing with Eqs.~(\ref{eq:ImA}) and (\ref{eq:optical}), we find
\begin{equation}
2\left(
\begin{array}{l}
\im\, A_{D^0\pi^-}\\
\im\, A_{D^+\pi^-}\\
\im\, A_{D^0\pi^0}
\end{array}
\right)
={\cal T}^\dagger
\left(
\begin{array}{l}
A_{D^0\pi^-}\\
A_{D^+\pi^-}\\
A_{D^0\pi^0}
\end{array}
\right),
\label{eq:2ImA=TA}
\end{equation}
\begin{equation}
{\cal T}^\dagger=\left(
\begin{array}{ccc}
r_0^*+r^*_e &0 &0\\
0 &r^*_0+r^*_a &{1\over \sqrt2}(r^*_a-r^*_e)\\
0 &{1\over \sqrt2}(r^*_a-r^*_e) &r^*_0+{1\over2}(r^*_a+r^*_e)\\
\end{array}
\right),
\label{eq:T}
\end{equation}
with ${\cal S}={\bold 1}+i{\cal T}$.
Eq.~(\ref{eq:2ImA=TA}) is consistent with Eq.~(\ref{eq:FSI}) through the identity 
$2\,\im\,{\cal S}^{1/2}={\cal T}^\dagger {\cal S}^{1/2}$ for symmetric ${\cal S}$. 
We also note that ${\cal T}$ can be diagonalized by using
${\cal T}=U^{\rm T}{\cal T}_{\rm diag}\,U$,
where $U$ is given in Eq.~(\ref{eq:U}),
giving
\begin{equation}
{\cal T}_{\rm diag}={\rm diag}\big(r_0+r_e,\;r_0+r_e,\;
r_0+{1\over2}(3r_a-r_e)\big).
\label{eq:TdiagSU2}
\end{equation}

\begin{figure}[t!]
\centerline{\hskip-1.1cm \DESepsf(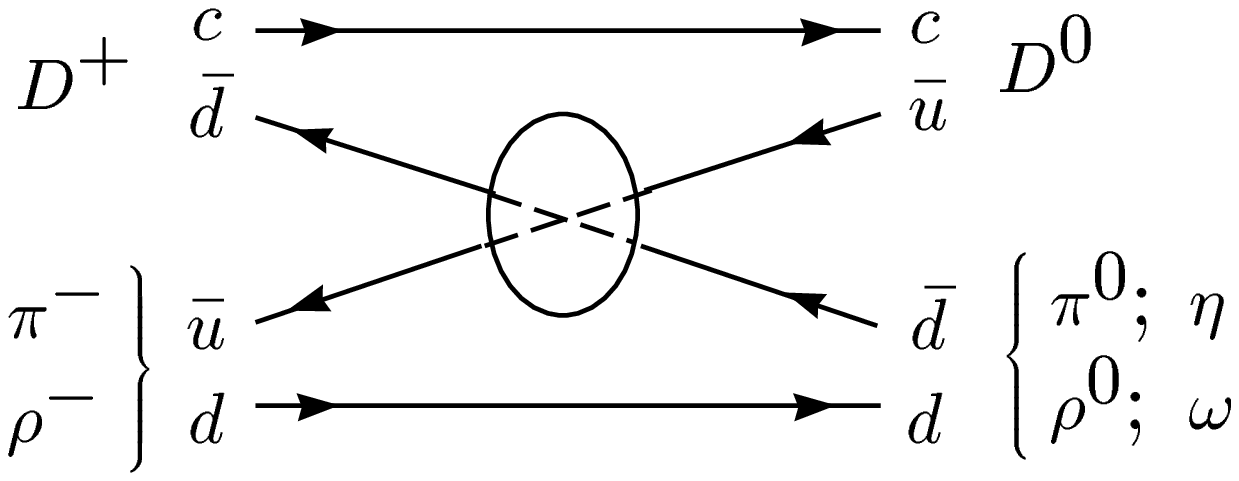 width 6.9cm)}
%
\smallskip
\centerline{\hskip0.5cm \DESepsf(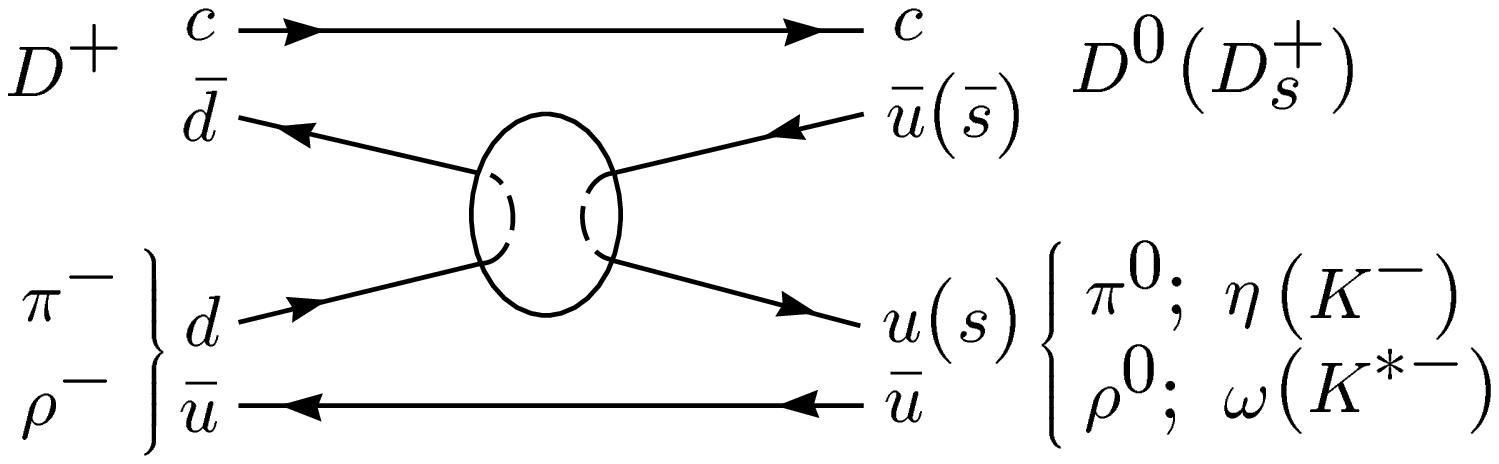 width 8.3cm)}
%
\smallskip
\centerline{\hskip-1.6cm \DESepsf(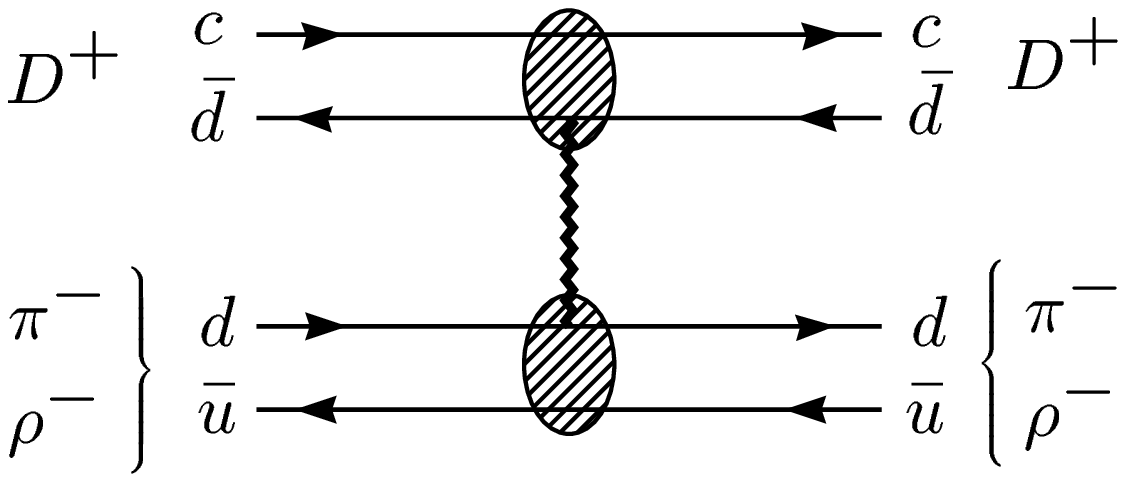 width 6.3cm)}
\smallskip
\caption{Pictorial representation of
  (a) $r_e$ (charge exchange), (b) $r_a$ (annihilation) and (c) $r_0$ (singlet exchange).
}
\label{fig:r2r1r0}
\end{figure}

The unitary scattering matrix ${\cal S}={\bf 1}+i {\cal T}$ can be
solved by identifying the elements in ${\cal T}_{\rm diag}$ with
$2\sin ({\rm angle})\, {\rm exp}(i\, {\rm angle})$'s, where we
note that $1+i\,2\sin ({\rm angle})\, {\rm exp}(i\, {\rm angle})=
{\rm exp}(i\,2{\rm angle})$.
One can now reproduce Eq.~(\ref{eq:S12}) by taking
\begin{eqnarray}
r_0+r_e&=&2\sin\delta_{3/2}\,e^{i\delta_{3/2}},
\nonumber\\
r_0+{1\over2}(3r_a-r_e)&=&2\sin\delta_{1/2}\,e^{i\delta_{1/2}}.
\label{eq:r1r2}
\end{eqnarray}
We give a pictorial representation of $r_e$, $r_a$, $r_0$ in Fig.~\ref{fig:r2r1r0};
they correspond to charge exchange, annihilation,
and flavor singlet exchange rescatterings, respectively.
Since the quark model is a representation of 
flavor SU(2) (SU(3)) group,
it should be able to reproduce the structure of ${\cal T}$, 
which follows from symmetry argument.
In particular the coefficients of $r_i$ in Eq. (\ref{eq:T})
can be reproduced easily using this pictorial approach
by matching the flavor wave function coefficients.
For example, we have $(r_a-r_e)/\sqrt2$ for a $D^+\pi^-$ to $D^0\pi^0$
rescattering.
We see from the first diagram of Fig. 4 that the exchange rescattering ($r_e$)
projects out the $d\bar d$ component of $\pi^0$ in the right hand side.
From our convention in Eq. (\ref{eq:isospinstructure}), this give a $-1/\sqrt2$
factor from the wave function of $\pi^0$.
Similarly the second diagram of Fig. 4 projects out the $u\bar u$ component 
of $\pi^0$ and give $r_a/\sqrt2$ consequently.
These diagrams also provide further information.
For example in the second diagram
as we go beyond SU(2), it is easy to see that
the annihilation rescattering ($r_a$)
is responsible for the $D^+\pi^-\to D^+_sK^-$ rescattering, 
since there is no $s$ quark before rescattering.

\section{Extension from SU(2) to SU(3)}

\subsection{$DP$ Modes }

We now generalize the $B\to D\pi$ case to SU(3) related modes
in the final state, such as $\overline B{}^0\to D^0\eta$ or $D^+_s K^-$.
We stress that we apply SU(3) symmetry only towards final state rescattering
rather than to the whole decay process. It is thus different from the
usual application of SU(3) in $B$ decays \cite{Grinstein:1996us},
where one decomposes the $B$ meson weak decay amplitudes,
including the effective Hamiltonian itself, into different SU(3) pieces,
and try to relate different modes (oftentimes including $B_s$ decay).
As argued earlier,
SU(3) should be a good symmetry for energetic FSI rescattering,
which is the case of interest. As we will see this approach give identical
results with the SU(3) decomposition approach shown in Appendix A.

It is straightforward to follow the steps through
Eqs.~(\ref{eq:operator})--(\ref{eq:r1r2}).
Eq.~(\ref{eq:operator}) remains unchanged, i.e.
\begin{eqnarray}
T&=&
\prod_{i,j=1,2} \int {d^3p_i\over (2\pi)^3 2 E_i}
                               {d^3q_j\over (2\pi)^3 2 E_j}
\nonumber\\
&\times& (2\pi)^4 \delta^4 (p_1+p_2-q_1-q_2)
\; {\cal D}^\dagger {\cal M} {\cal D}^\prime,
\label{eq:operatorSU3}
\end{eqnarray}
but now the multiplets are extended
\begin{eqnarray}
{\cal D}^\dagger&=&\big(a_{D^0}({p_1})\,\ a_{D^+}({p_1})\,\ a_{D^+_s}({p_1})\big),
\nonumber\\
{\cal D}^{\prime\rm T}\!\!&=&
\big(a^\dagger_{D^0}({q_1})\,\ a^\dagger_{D^+}({q_1})\,\ a^\dagger_{D^+_s}({q_1})\big),
\nonumber\\
{\cal M}&=&M_0(q_1q_2;p_1p_2)\,{\rm Tr}\big(\overline\Pi\,\Pi\big)\,{\bf 1}
\nonumber\\
&&
+M_a(q_1q_2;p_1p_2)\,\Pi\,\overline\Pi
+M_e(q_1q_2;p_1p_2)\,\overline\Pi\,\Pi,
\nonumber \\
\Pi({p_2})&=&
\nonumber\\
&&\hspace{-1cm}
\left(
\begin{array}{ccc}
{a_{\pi^0}\over\sqrt2}+{a_{\eta_8}\over\sqrt6} &a_{\pi^+}
&a_{K^+}\\
a_{\pi^-} &-{a_{\pi^0}\over\sqrt2}+{a_{\eta_8}\over\sqrt6}
&a_{K^0}\\
a_{K^-}  &a_{\overline K{}^0}          &-\sqrt{2\over3} a_{\eta_8}
\end{array}
\right)({p_2}),
\nonumber \\
\overline \Pi({q_2})&=&
\nonumber\\
&&\hspace{-1cm}
\left(
\begin{array}{ccc}
{a^\dagger_{\pi^0}\over\sqrt2}+{a^\dagger_{\eta_8}\over\sqrt6}
&a^\dagger_{\pi^-} &a^\dagger_{K^-}\\
a^\dagger_{\pi^+}
&-{a^\dagger_{\pi^0}\over\sqrt2}+{a^\dagger_{\eta_8}\over\sqrt6}
&a^\dagger_{\overline K{}^0}\\
a^\dagger_{K^+}  &a^\dagger_{K^0}
&-\sqrt{2\over3}a^\dagger_{\eta_8}
\end{array}
\right)({q_2}).
\label{eq:pipibarSU3}
\end{eqnarray}
Note that this operator can rescatter $D^+\pi^-,\,D^0\pi^0$
into the desired states $D^0\eta_8,\,D^+_s K^-$.

The physical $\eta,\,\eta^\prime$ mesons are defined through
\begin{equation}
\left(
\begin{array}{c}
\eta\\
      \eta^\prime
\end{array}
\right)=
\left(
\begin{array}{cc}
\cos\vartheta &-\sin\vartheta\\
\sin\vartheta &\cos\vartheta
\end{array}
\right)
\left(
\begin{array}{c}
\eta_8\\
      \eta_1
\end{array}
\right),
\end{equation}
where the mixing angle $\vartheta=-15.4^\circ$ \cite{Feldmann:1998vh}.
In principle, we should also include $D^0\eta_1$ in the rescattering process.
The additional terms can be obtained by
replacing $\Pi$ in Eq.~(\ref{eq:pipibarSU3}) by
$\Pi+{\bf 1}\,a_{\eta_1}(q_2)/\sqrt3$ (and similarly for $\overline\Pi$),
and labeling the $\eta_1$ related matrix elements by
$M^\prime_i(q_1q_2;p_1p_2)$.
Knowing that the U$_A(1)$ symmetry is
broken by anomaly and $\eta_1$ is not a Goldstone boson,
$M^\prime_i$ are not identical to $M_i$.
The number of parameters would therefore double,
but experimental measurements are still scarce.
On the other hand, we note that the mixing angle $\vartheta$ is quite small,
so we approximate $\eta$ by $\eta_8$.
Thus,
we concentrate on the rescattering process involving
octet pseudoscalar mesons only, as a step beyond the elastic
FSI discussed in the previous section.
In this way, as already shown, one again has just three independent amplitudes.

Besides Eq.~(\ref{eq:M1M2}) we now have
\begin{eqnarray}
M(D^+\pi^- \to D^0{\eta_8})
&=&{1\over\sqrt6}(M_a+M_e),
\nonumber\\
M(D^+\pi^- \to D^+_s K^-)&=&M_a,
\nonumber \\
M(D^0\pi^0 \to D^0{\eta_8})
&=&{1\over2\sqrt3}(M_a+M_e),
\nonumber\\
M(D^0\pi^0  \to D^+_s K^-)&=&{1\over\sqrt2}M_a,
\nonumber \\
M(D^0{\eta_8} \to D^0{\eta_8})&=&M_0
 + {1\over6}(M_a+M_e),
\nonumber \\
M(D^0{\eta_8}  \to D^+_s K^-)
&=&{1\over\sqrt6}(M_a-2 M_e),
\nonumber\\
M(D^+_s K^-  \to D^+_s K^-)
&=&M_0+M_a,
\label{eq:M1M2SU3}
\end{eqnarray}
where $M_i$ stands for $M_i(q_1q_2;p_1p_2)$.
Amplitudes for other related modes can be obtained by noting that
$M(ab\to cd)=M(cd\to ab)$ in our case.

With Eqs.~(\ref{eq:M1M2}), (\ref{eq:M1M2SU3}) and similar
extensions of Eqs.~(\ref{eq:imArA}) and (\ref{eq:r1r2def}),
we extend Eqs.~(\ref{eq:2ImA=TA}) and (\ref{eq:T}) to
$2\,\im\,A={\cal T}^\dagger\,A$ in the
$D^0\pi^-$, $D^+\pi^-$, $D^0\pi^0$, $D^0\eta_8$ and $D^+_s K^-$
basis with
\begin{equation}
{\cal T}=r_0{\bf 1} +
\left(
\begin{array}{ccccc}
r_e & 0 & 0 & 0 & 0 \\
0 & r_a
  & {r_a-r_e\over\sqrt2}
  & {r_a+r_e\over\sqrt6}
  & r_a \\
0 & {r_a-r_e\over\sqrt2}
  & {r_a+r_e\over2}
  & {r_a+r_e\over2\sqrt3}
  & {r_a\over\sqrt2} \\
0 & {r_a+r_e\over\sqrt6}
  & {r_a+r_e\over2\sqrt3}
  & {r_a+r_e\over6}
  & {r_a-2r_e\over\sqrt6} \\
0 & r_a
  & {r_a\over\sqrt2}
  & {r_a-2r_e\over\sqrt6}
  & r_a
\end{array}
\right),
\label{eq:TSU3}
\end{equation}
where ${\cal T}$ can also be easily obtained by the pictorial approach,
as we explain in the end of the previous section.
Note that ${\cal T}$ can be diagonalized as
\begin{eqnarray}
{\cal T}_{\rm diag} =r_0\,{\bf 1}
 + {\rm diag}\left(r_e,\,r_e,\,r_e,\,-r_e,\,{1\over3}(8r_a-r_e)\right).
\end{eqnarray}

Following similar procedure in the previous section,
${\cal S}=1+i{\cal T}$ is obtained by the (physical) substitution
\begin{eqnarray}
(1+i r_0)\, e^{-2i\delta_{3/2}}&=&{1\over2}\,
          (1+e^{2i\delta^\prime}),
\nonumber\\
i r_e\, e^{-2i\delta_{3/2}}&=&{1\over2}\,
      (1-e^{2i\delta^\prime}),
\nonumber\\
i r_a\, e^{-2i\delta_{3/2}}&=&{1\over8}\,
 (-1-2\,e^{2i\delta^\prime}+3\,e^{2i\theta}),
\label{eq:r1r2SU3}
\end{eqnarray}
where $r_0+r_e = 2\sin\delta_{3/2}\,e^{i\delta_{3/2}}$ as in Eq. (\ref{eq:r1r2}), 
but $\delta = \delta_{1/2} - \delta_{3/2}$ is now extended to two physical phase 
differences $\delta^\prime$ and $\theta$.
Note that $D$, $D_s$ and $\pi$, $K$, $\eta$ transform as $\overline 3$ and $8$ 
under SU(3). 
As shown in Appendix A, we can identify above phases as
\begin{equation}
\delta_{3/2}=\delta_{\overline{15}},\quad 
\delta^\prime=\delta_6-\delta_{\overline{15}},\quad 
\theta=\delta_{\overline 3}-\delta_{\overline{15}}.   
\end{equation}

By analogy to the previous section,
the solution of $2\,\im\,A={\cal T}^\dagger A$ is  
\begin{equation}
\left(
\begin{array}{l}
A_{D^0\pi^-}\\
A_{D^+\pi^-}\\
A_{D^0\pi^0}\\
A_{D^0\eta_8}\\
A_{D^+_s K^-}
\end{array}
\right)
={\cal S}^{1/2}\,
\left(
\begin{array}{l}
A^f_{D^0\pi^-}\\
A^f_{D^+\pi^-}\\
A^f_{D^0\pi^0}\\
A^f_{D^0\eta_8}\\
A^f_{D^+_s K^-}
\end{array}
\right),
 \label{eq:FSISU3}
\end{equation}
where $A^f$s are factorization amplitudes.
The matrix ${\cal S}_{1/2}$ can be obtained
by reducing phases in ${\cal S}$ by half.
For later purpose, we give the explicit expression of ${\cal S}^{1/2}$
(or equivalently ${\cal S}$ with trivial modification of phases),
\begin{widetext}
\begin{eqnarray}
&& {\cal S}^{1/2}e^{-i\delta_{3/2}}
\nonumber\\
 &=&  \left(
\begin{array}{ccccc}
1 & 0 & 0 & 0 & 0 \\
0 & {1\over8}(3+2e^{i\delta^\prime}+3e^{i\theta})
  & {1\over8\sqrt2}(-5+2e^{i\delta^\prime}+3e^{i\theta})
  & {1\over8}\sqrt{3\over2}(1-2e^{i\delta^\prime}+e^{i\theta})
  &-{1\over8}(1+2e^{i\delta^\prime}-3e^{i\theta}) \\
0 & {1\over8\sqrt2}(-5+2e^{i\delta^\prime}+3e^{i\theta})
  & {1\over16}(11+2e^{i\delta^\prime}+3e^{i\theta})
  & {\sqrt3\over16}(1-2e^{i\delta^\prime}+e^{i\theta})
  &-{1\over8\sqrt2}(1+2e^{i\delta^\prime}-3e^{i\theta}) \\
0 & {1\over8}\sqrt{3\over2}(1-2e^{i\delta^\prime}+e^{i\theta})
  & {\sqrt3\over16}(1-2e^{i\delta^\prime}+e^{i\theta})
  & {1\over16}(9+6e^{i\delta^\prime}+e^{i\theta})
  & {1\over8}\sqrt{3\over2}(-3+2e^{i\delta^\prime}+e^{i\theta}) \\
0 & -{1\over8}(1+2e^{i\delta^\prime}-3e^{i\theta})
  & -{1\over8\sqrt2}(1+2e^{i\delta^\prime}-3e^{i\theta})
  & {1\over8}\sqrt{3\over2}(-3+2e^{i\delta^\prime}+e^{i\theta})
  & {1\over8}(3+2e^{i\delta^\prime}+3e^{i\theta})
\end{array}
\right).
\label{eq:S12SU3}
\end{eqnarray}
\end{widetext}
Just as in Eqs.~(\ref{eq:FSI}) and (\ref{eq:S12}),
${\cal S}^{1/2}$ of Eq. (\ref{eq:S12SU3}) has an overall phase.
Only phase differences affect decay rates.
An overall sign change of the phases also leaves rates unchanged.

Note that charge conservation and unitarity imply,
\begin{eqnarray}
&&|A_{D^0\pi^-}|^2=|A^f_{D^0\pi^-}|^2,
\nonumber\\
&&|A_{D^+\pi^-}|^2+|A_{D^0\pi^0}|^2+|A_{D^0\eta}|^2+|A_{D^+_s K^-}|^2
\nonumber\\
&=&
|A^f_{D^+\pi^-}|^2+|A^f_{D^0\pi^0}|^2+|A^f_{D^0\eta}|^2+|A^f_{D^+_s K^-}|^2.
\qquad\label{eq:unitarySU3}
\end{eqnarray}
Since the amplitudes for color suppressed modes are small 
at the factorization level,
they will be fed by the color allowed amplitude $A^f_{D^+\pi^-}$.
As a consequence, the $D^+\pi^-$ rate will be reduced from its factorization result.
Compared with the elastic case, we now have additional, but slight, 
leakage of $D^+\pi^-$ into $D^0\eta$ and $D^+_s K^-$ modes.
Because the measured rates of color suppressed modes
are still small as compared to the color allowed modes,
as far as the $D\pi$ system is concerned, 
the results do not deviate too much from the previous section.

On the other hand,
since the factorization amplitudes $A^f_{D\pi}$ satisfy
the isospin triangular relation, Eq.~(\ref{eq:triangle}),
one can show that, with FSI
and for any value of $A^f_{D^0\eta_8}$ and $A^f_{D^+_s K^-}$,
the rescattered $A_{D\pi}$ amplitudes also satisfy the relation.
As noted earlier,
the isospin relation should hold whether FSI is active or not.
To demonstrate this,
it is instructive to express the FSI of Eq.~(\ref{eq:S12SU3})
in the isospin basis.
In the
$D^0\pi^-$, $(D\pi)_{3/2}$, $(D\pi)_{1/2}$, $D^0\eta_8$ and $D^+_s K^-$
basis, the ${\cal T}$ matrix, and similarly ${\cal S}^{(1/2)}$,
take a block diagonal form 
\begin{equation}
{\cal T}_{\rm block}=r_0{\bf 1}
+\left(
\begin{array}{ccccc}
r_e & 0 & 0 & 0 & 0 \\
0 & r_e & 0 & 0 & 0 \\
0 & 0
  & {3r_a-r_e\over2}
  & {r_a+r_e\over2}
  & \sqrt{3\over2} r_a \\
0 & 0
  & {r_a+r_e\over2}
  & {r_a+r_e\over6}
  & {r_a-2r_e\over\sqrt6} \\
0 & 0
  & \sqrt{3\over2} r_a
  & {r_a-2r_e\over\sqrt6}
  & r_a
\end{array}
\right),
\label{eq:Tblock}
\end{equation}
by
${\cal T}_{\rm block}=O\,{\cal T}\,O^{\rm T}$,
where
\begin{equation}
O=
\left(
\begin{array}{ccc}
U &0 &0 \\
0 &1 &0\\
0 &0 &1
\end{array}
\right),
\end{equation}
and $U$ is given in Eq.~(\ref{eq:U}). 
Truncating to the first $3 \times 3$ sub-matrix,
${\cal T}$ is diagonal and one reproduces
the SU(2) case of Eq. (\ref{eq:TdiagSU2}).

In this basis, with block
diagonalized ${\cal S}^{1/2}_{\rm block}=O\,{\cal S}^{1/2}\,O^{\rm
T}$, Eq.~(\ref{eq:FSISU3}) becomes
\begin{equation}
A={\cal S}^{1/2}_{\rm block}\,A^f,
\label{eq:FSIblock}
\end{equation}
where the first two diagonal elements of ${\cal S}^{1/2}_{\rm block}$
are just $e^{i\delta_{3/2}}$, while the remaining lower block governs
the ``inelastic'' rescatterings
\begin{eqnarray}
e^{-i\delta_{3/2}}\, ({\cal S}^{1/2}_{\rm block})_{(D\pi)_{1/2},(D\pi)_{1/2}}
&=&{1\over16}(1+6e^{i\delta^\prime}\!\!\!+9e^{i\theta}),
\nonumber\\
e^{-i\delta_{3/2}} \, ({\cal S}^{1/2}_{\rm block})_{(D\pi)_{1/2},D^0\eta_8}
&=&{3\over16}(1-2e^{i\delta^\prime}\!\!\!+e^{i\theta}),
\nonumber \\
e^{-i\delta_{3/2}} \, ({\cal S}^{1/2}_{\rm block})_{(D\pi)_{1/2},D^+_s K^-}
&=&-{\sqrt{3}\over8\sqrt2}(1+2e^{i\delta^\prime}\!\!\!-3e^{i\theta}),
\nonumber\\
e^{-i\delta_{3/2}} \, ({\cal S}^{1/2}_{\rm block})_{D^0\eta_8,D^0\eta_8}
&=&{1\over16}(9+6e^{i\delta^\prime}\!\!\!+e^{i\theta}),
\nonumber\\
e^{-i\delta_{3/2}} \, ({\cal S}^{1/2}_{\rm block})_{D^0\eta_8,D^+_sK^-}
&=&{\sqrt{3}\over8\sqrt2}(-3+2e^{i\delta^\prime}\!\!\!+e^{i\theta}),
\nonumber\\
e^{-i\delta_{3/2}} \, ({\cal S}^{1/2}_{\rm block})_{D^+_sK^-,D^+_sK^-}
&=&{1\over8}(3+2e^{i\delta^\prime}\!\!\!+3e^{i\theta}).
\label{eq:S12block}
\end{eqnarray}
between the $I=I_z=1/2$ decay final states $(D\pi)_{1/2}$, $D^0\eta_8$ and $D^+_s K^-$,
which is a reasonable extension beyond the elastic rescattering 
discussed in the previous section.
The elastic case corresponds to 
$e^{-i\delta_{3/2}} ({\cal S}^{1/2}_{\rm block})_{(D\pi)_{1/2},(D\pi)_{1/2}}
=e^{i\delta}$ and setting the rest of Eq. (\ref{eq:S12block}) to zero.
Since the factorization amplitudes for $D^0\eta$ and $D^+_s K^-$ are small,
their FSI contribution to the $D\pi$ system will be suppressed.
Therefore, 
\begin{eqnarray}
\left|{A_{1/2}\over A^f_{1/2}}\right|
&\sim& \left|{1\over16}(1+6e^{i\delta^\prime}\!\!\!+9e^{i\theta})\right|,
\nonumber\\
\delta
&\sim&\arg\,(1+6e^{i\delta^\prime}\!\!\!+9e^{i\theta}),
\label{eq:estimate}
\end{eqnarray}
are good estimates.
The geometric meaning of Eq. (\ref{eq:S12block}) is given in Appendix B.

\subsection{$D^{*}P$ and $DV$ Modes}

The formalism can be applied to final states involving
pseudoscalar and vector mesons (PV), with only slight modifications.

For $D^*(p_1;\lambda) P(p_2)\to
D^*(q_1;\lambda^\prime) P^\prime(q_2)$ rescattering, where
$\lambda^{(\prime)}$ is the polarization index and $P^{(\prime)}$
denotes a pseudoscalar meson, we replace $a_D(p_1)$ and
$M_i(q_1q_2;p_1p_2)$ by $a_{D^*}(p_1,\lambda)$ and
$M_i(q_1q_2,\lambda^\prime;p_1p_2,\lambda)$ etc., respectively,
where
\begin{eqnarray}
&&\!\!\!\!\!\!\!\!\!\!\!\!\!\!\!\!\! \langle
q_1q_2,\lambda^\prime| T |p_1p_2,\lambda\rangle
\nonumber \\
&&\!\!\!\!\!\!\!\!\!\!\!\!= (2\pi)^4\delta^4(p_1+p_2-q_1-q_2)
M(q_1q_2,\lambda^\prime;p_1p_2,\lambda) .
\end{eqnarray}
The $B\to D^*P$ amplitude is expressed as
\begin{equation}
M(B\to D^*P)=\varepsilon_\lambda^*\cdot p_B \, A_{D^*P},
 \label{eq:MVP}
\end{equation}
where $\varepsilon_\lambda$ is the polarization vector and
$A_{D^*P}$ is a Lorentz scalar that is independent of
$\varepsilon$ and the angle between the 3-momenta of $D^*$ and $P$.

Take $B^-\to D^{*0}\pi^-$ for example. 
Eq.~(\ref{eq:imArA}) becomes
\begin{eqnarray}
&& 2\,\im\,[(\varepsilon^*_{\lambda}\cdot p_B)
A_{D^{*0}\pi^-}]
\nonumber \\
&=&\sum_{\lambda^\prime}\int {d^3q_1\over (2\pi)^3 2 E_1}{d^3q_2\over
(2\pi)^3 2 E_2}
                                    (2\pi)^4 \delta^4 (p_1+p_2-q_1-q_2)
\nonumber \\
&& \times \big[M^*_0(q_1q_2,\lambda^\prime;p_1p_2,\lambda)
             +M^*_e(q_1q_2,\lambda^\prime;p_1p_2,\lambda)\big]
\nonumber \\
&& \times     \varepsilon^*_{\lambda^\prime}\cdot p_B \,
A_{D^{*0}\pi^-}.
\end{eqnarray}
By choosing a real basis for $\varepsilon$ and noting that
$\sum_\lambda (p_B\cdot \varepsilon_\lambda)^2=p_{\rm
cm}^2\,m_B^2/m_{D^{*0}}^2 $, 
where $p_{\rm cm}$ is the momentum of $D^{*0}$ in the center of mass frame,
we obtain
\begin{equation}
2\,\im\,A_{D^{*0}\pi^-}=(r_0^*+r_e^*) \, A_{D^{*0}\pi^-},
\end{equation}
with
\begin{eqnarray}
r^*_{i}&\equiv& \sum_{\lambda\lambda^\prime} \int {d^3q_1\over
(2\pi)^3 2 E_1}{d^3q_2\over (2\pi)^3 2 E_2}
\nonumber \\
&&\times(2\pi)^4\delta^4 (p_1+p_2-q_1-q_2)\, {m^2_{D^{*0}}\over
m_B^2 p_{cm}^2}
\nonumber\\
&&\times \varepsilon_\lambda\cdot p_B \,
M^*_{i}(q_1q_2\lambda^\prime;p_1p_2,\lambda) \,
\varepsilon_{\lambda^\prime}\cdot p_B,
\label{eq:r1r2defVP}
\end{eqnarray}
which projects out the P-wave $D^*P$ scattering amplitude.
Equations after Eq. (\ref{eq:pipibarSU3}) from previous section
can be carried over by replacing $A_{DP}\to A_{D^*P}$.

The generalization to $DV$ decay modes is again straightforward, 
except that $\omega_1$--$\omega_8$ mixing cannot be neglected.
The physical mesons should be
\begin{eqnarray}
\omega&=&\sqrt{2\over3}\omega_1+\sqrt{1\over3}\omega_8,
\nonumber \\
\phi&=&\sqrt{1\over3}\omega_1-\sqrt{2\over3}\omega_8.
\label{eq:omegaphi}
\end{eqnarray}
Unlike the pseudoscalar case, where ignoring $\eta_1$ couplings
can be partially justified, $\omega_1$ in principle should be on
similar footing as $\omega_8$.

We replace $\pi$, $K$ and $\eta_8$ in Eq. (\ref{eq:pipibarSU3})
by $\rho$, $K^*$ and $\omega_8$.
By including $\omega_1$ we will have two more terms in ${\cal M}$ of
the corresponding $T$ matrix namely,
$a^\dagger_{\omega_1}(q_2) a_{\omega_1}(p_2)
[\widetilde M_0+(\widetilde M_a+\widetilde M_e)/3]$ and
$[a_{\omega_1}(p_2)\,\bar\Pi+\Pi\,a^\dagger_{\omega_1}(q_2)]
 (\overline M_a+\overline M_e)/\sqrt3$.
These two terms correspond to $\omega_1$ to $\omega_1$ and 
$\omega_1$ to octet rescattering. 
These $\widetilde M_i$, $\overline M_i$ will reduce to $M_i$ under nonet
symmetry.   


The $\cal T$ matrix can be obtained as before.
In isospin basis with the 
$D^0\rho^-$, $(D\rho)_{3/2}$, $(D\rho)_{1/2}$, $D^0\omega_8$, $D^+_sK^{*-}$ and
$D^0\omega_1$, we have 
\begin{eqnarray}
{\cal T}_{\rm block}&=&{\rm diag}
\left({r_0,\,r_0,\,r_0,\,r_0,\,r_0,\,\widetilde r_0}\right)
\nonumber\\
&&\hspace{-1cm}
+\left(
\begin{array}{cccccc}
r_e & 0 & 0 & 0 & 0 &0\\
0 & r_e & 0 & 0 & 0 &0\\
0 & 0
  & {3r_a-r_e\over2}
  & {r_a+r_e\over2}
  & \sqrt{3\over2} r_a 
  & {\bar r_a+\bar r_e\over\sqrt2} \\
0 & 0
  & {r_a+r_e\over2}
  & {r_a+r_e\over6}
  & {r_a-2r_e\over\sqrt6} 
  & {\bar r_a+\bar r_e\over3\sqrt2}\\
0 & 0
  & \sqrt{3\over2} r_a
  & {r_a-2r_e\over\sqrt6}
  & r_a
  & {\bar r_a+\bar r_e\over\sqrt3}\\
0 & 0
  & {\bar r_a+\bar r_e\over\sqrt2}
  & {\bar r_a+\bar r_e\over3\sqrt2}
  & {\bar r_a+\bar r_e\over\sqrt3}
  & {\widetilde r_a+\widetilde r_e\over3}
\end{array}
\right).
\label{eq:TblockV}
\end{eqnarray}
It can also be easily obtained by the pictorial approach in the $\omega_1$,
$\omega_8$ basis.
${\cal T}_{\rm block}$ is identical to Eq. (\ref{eq:Tblock}) except the 
additional sixth row and column.
The solution is, $r_{0,e}$ as shown in Eq. (\ref{eq:r1r2SU3}) and
\begin{eqnarray}
i r_a\,e^{-2i \delta_{3/2}}&=&{1\over 8}(3\,{\cal U}_{\overline 3\overline 3} 
                       e^{-2i \delta_{3/2}}-2 e^{2 i\delta^\prime}-1),
\nonumber\\
i (\bar r_a+\bar r_e)&=&
{3\over2\sqrt2} {\cal U}_{\overline 3\overline 3{}^\prime},
\nonumber \\
i( \widetilde r_0+{\widetilde r_a+\widetilde r_e\over3})&=&
{\cal U}_{\overline 3{}^\prime\overline 3{}^\prime}-1,
\end{eqnarray}
where ${\cal U}$ is a two by two symmetric unitary matrix
that mixes $\overline{\bold 3}$ and $\overline{\bold 3}^\prime$ 
by rescattering as shown in Appendix A.
Note that we need two phases and one mixing angle to specify ${\cal U}$,
resulting four parameters (with an overall phase factored out) to describe
the rescattering matrix.
There are too many parameters and experiment measurements are still scare.
However, there is no U$_A$(1) anomaly in the vector sector,
and we expect U(3) rather than SU(3) symmetry.
Therefore we consider rescattering among $D^{0,+},D_s^+$
and the vector nonet as a first step
beyond the simple elastic FSI case.
In this case we identify $\bar r_i$ and $\widetilde r_i$ as $r_i$.

It turns out that U(3) symmetry allows only
either charge exchange or annihilation FSI, but not both.
The two distinct solutions for $r_i$ and $\cal U$ are
\vskip0.2cm\noindent
\underline{Solution 1}:
\begin{eqnarray}
(1+i r_0) \, e^{-2i\delta_{3/2}} &=& {1\over2}\,(1+e^{2i\delta^\prime}),
\nonumber\\
i r_e \, e^{-2i\delta_{3/2}} &=& {1\over2}\,(1-e^{2i\delta^\prime}),
\nonumber\\
i r_a \, e^{-2i\delta_{3/2}} &=& 0,
\nonumber\\
{\cal U}_{\overline 3\overline3} e^{-2i\delta_{3/2}}&=&{1\over3}(1+2 e^{2i\delta^\prime}),
\nonumber\\
{\cal U}_{\overline 3\overline3{}^\prime} e^{-2i\delta_{3/2}}&=&{\sqrt2\over3}(1-e^{2i\delta^\prime}),
\nonumber\\
{\cal U}_{\overline 3{}^\prime\overline3{}^\prime} e^{-2i\delta_{3/2}}&=&{1\over3}(2+ e^{2i\delta^\prime}),
\label{eq:Sol1}
\end{eqnarray}
\underline{Solution 2}:
\begin{eqnarray}
(1+i r^{(\prime)}_0) \, e^{-(2)i\delta_{3/2}} &=& 1,
\nonumber\\
i r^{(\prime)}_e \, e^{-(2)i\delta_{3/2}} &=& 0.
\nonumber\\
i r^{(\prime)}_a \, e^{-(2)i\delta_{3/2}} &=& {1\over3}\,(-1+e^{(2)i\sigma}),
\nonumber\\
{\cal U}_{\overline 3\overline3} e^{-2i\delta_{3/2}}&=&{1\over9}(1+ 8e^{2i\sigma}),
\nonumber\\
{\cal U}_{\overline 3\overline3{}^\prime} e^{-2i\delta_{3/2}}&=&{2\sqrt2\over9}(-1+e^{2i\sigma}),
\nonumber\\
{\cal U}_{\overline 3{}^\prime\overline3{}^\prime} e^{-2i\delta_{3/2}}&=&{1\over9}(8+ e^{2i\sigma}),
\label{eq:Sol2}
\end{eqnarray}
where the former (latter) does not have annihilation (exchange) contribution.
Notice that while $\delta^\prime\equiv\delta_6-\delta_{\overline{15}}$ is analogous to the $D^{(*)}P$ counterpart,
$\sigma$ is from $\cal U$ and is not equivalent to $\theta$.

To understand why $r_a r_e=0$, we shown the ${\cal T}$ matrix in the
$D^0\rho^-$, $D^+\rho^-$, $D^0\rho^0$, $D^0\omega$, $D^+_s K^{*-}$ and $D^0\phi$
basis.
\begin{equation}
{\cal T}_V=r_0{\bf 1}
+\left(
\begin{array}{cccccc}
r_e & 0 & 0 & 0 & 0 & 0 \\
0 & r_a
  & {r_a-r_e\over\sqrt2}
  & {r_a+r_e\over\sqrt2}
  & r_a
  & 0 \\
0 & {r_a-r_e\over\sqrt2}
  & {r_a+r_e\over2}
  & {r_a+r_e\over2}
  & {r_a\over\sqrt2}
  & 0 \\
0 & {r_a+r_e\over\sqrt2}
  & {r_a+r_e\over2}
  & {r_a+r_e\over2}
  & {r_a\over\sqrt2}
  & 0 \\
0 & r_a
  & {r_a\over\sqrt2}
  & {r_a\over\sqrt2}
  & r_a
  & r_e \\
0 & 0
  & 0
  & 0
  & r_e
  & 0 \\
\end{array}
\right).
\label{eq:Tv}
\end{equation}
Note that ${\cal T}_V^{6i}=0$ for $i=2,3,4$, 
so $D^0\phi$ can only rescatter with itself and $D^+_s K^{*-}$.
This can be easily checked by using the pictorial method as shown in Fig.~4.
 
%
%
We define ${\cal T}^\prime_V$ via
${\cal S}^{1/2}_V=1+i {\cal T}^\prime_V$.
It is easy to show that 
${\cal T}_V=2{\cal T}^\prime_V+i {\cal T}_V^{\prime 2}$.  
${\cal S}^{1/2}_V$ should also satisfy the U(3) symmetry as well
since it is generated by the same dynamics (or Hamiltonian) as
${\cal S}_V$.
Since the construction of ${\cal T}_V$ is based on symmetry,
we expect ${\cal T}^\prime_V$ to have the same structure,
or simply with $r_{0,a,e}$ replaced by $r_{0,a,e}^\prime$.
It is then easy to show that 
\begin{equation}
{\cal T}^{6j}_V=i({\cal T}_V^{\prime 2})^{6j}\propto r_a^\prime\,r_e^\prime,
\end{equation}
for $j=2,3,4$.
Since these elements of ${\cal T}_V$ are zero,
we must have $r^\prime_a\,r^\prime_e=0$
which implies $r_a\,r_e=0$.

With these $r^{\prime}_i$, which are $r_i$ in 
Eqs. (\ref{eq:Sol1}), (\ref{eq:Sol2}) with phases reduced by half,
we obtain the decay amplitudes by applying
$A={\cal S}^{1/2}_V\,A^f$.

\section{\label{sec:num} Numerical analysis}

\subsection{$\overline B{}\to DP$ and $D^*P$ Modes}

For $\overline B\to D\pi$ modes, we start from Eq.~(\ref{eq:Af}),
where the factorization amplitudes are decomposed into
color-allowed external $W$-emission (${\it T}$),
color-suppressed internal $W$-emission (${\it C}$),
and $W$-exchange amplitude (${\it E}$).
They are given by \cite{Neubert:1997uc,Cheng:1999kd}
\begin{eqnarray}
{\it T}&=&{G_F\over\sqrt2} V_{cb} V_{ud}^* \, a_1 \, (m_B^2-m_D^2) f_\pi
           F_0^{BD}(m^2_\pi),
\nonumber\\
{\it C}&=&{G_F\over\sqrt2} V_{cb} V_{ud}^* \, a_2 \, (m_B^2-m_\pi^2) f_D
           F_0^{B\pi}(m^2_D), \\
{\it E}&=&{G_F\over\sqrt2} V_{cb} V_{ud}^* \, a_2 \, (m_D^2-m_\pi^2) f_B
           F_0^{0\to D\pi}(m^2_B).
\nonumber
\end{eqnarray}
Since the annihilation form factor $F_0^{0\to D\pi}(m^2_B)$
is expected to be suppressed at $q^2=m_B^2$ and $a_2$ is small,
the amplitude $\it E$ is neglected.
For $\overline B{}^0\to D^0\eta$ we have
\begin{equation}
A^f_{D^0\eta}={G_F\over\sqrt2} V_{cb} V_{ud}^* a_2
(m_B^2-m_\eta^2)
              f_D F_0^{B\eta}(m^2_D),
\end{equation}
where $\eta_1$--$\eta_8$ mixing effect is included
via $F_0^{B\eta}(m^2_D)=\cos\vartheta\,
F_0^{B\eta_8}-\sin\vartheta\, F_0^{B\eta_1}$.
We use experimentally measured masses in
$A^f_{D^0\pi^-},\,A^f_{D^+\pi^-},\,A^f_{D^0\pi^0}$ and $A^f_{D^0\eta}$.
These amplitudes are real in our phase convention.

For $\overline B\to D^*\pi,\,D^{*0}\eta$, we have 
$M(\overline B\to D^* P)=(\varepsilon\cdot p_B) \, A_{D^*P}$ [Eq.~(\ref{eq:MVP})].
Analogous to Eq.~(\ref{eq:Af}),
the factorization amplitudes $A^f_{D^*\pi}$
are decomposed into
\begin{eqnarray}
{\it T}&=& {G_F\over\sqrt2}V_{cb} V_{ud}^*
	\, a_1 \, f_\pi 2m_{D^*} A^{BD}_0(m_\pi^2),
\nonumber \\
{\it C}&=& {G_F\over\sqrt2}V_{cb} V_{ud}^*
	\, a_2 \, f_{D^*} 2m_{D^*} F^{B\pi}_1(m_D^2),
\end{eqnarray}
where ${\it E}$ is again neglected.
For $\overline B{}^0\to D^{*0}\eta$ we have
\begin{equation}
A^f_{D^{*0}\eta}=
            {G_F\over\sqrt2}V_{cb} V_{ud}^*
	    a_2 f_{D^*} 2m_{D^*} F^{B\eta}_1(m_D^2).
\end{equation}

\begin{table}[t!]
\caption{
\label{tab:formfactor}
Form factors in LF and NS
form-factor models where
$A_i^{B\omega}(q^2)=A_i^{B\rho}(q^2)/\sqrt{2}$ and
$V^{B\omega}(q^2)=V^{B\rho}(q^2)/\sqrt{2}$.}
\begin{ruledtabular}
\begin{tabular}{lclc}
      & LF (NS)  & &LF (NS)\\
\hline
$F_0^{B\pi}(m_D^2)$ & 0.29 (0.27) &$A_0^{B{D^*}}(m_\pi^2)$ & 0.73 (0.64)\\
$F_1^{B\pi}(m^2_{D^*})$ & 0.34 (0.32)&$A_0^{B\rho}(m_D^2)$ & 0.35 (0.31)\\
$F_0^{B\eta}(m_D^2)$ & 0.16 (0.15)&$A_1^{B\rho}(m_{D^*}^2)$ & 0.23 (0.28)\\
$F_1^{B\eta}(m_{D^*}^2)$ & 0.19 (0.18)&$A_2^{B\rho}(m_{D^*}^2)$ &0.22 (0.31)\\
$F_0^{BD}(m_\pi^2)$ & 0.70 (0.63)&$V^{B\rho}(m_{D^*}^2)$ & 0.30 (0.32)\\
$F^{BD}_0(m^2_K)$ &0.70 (0.64)&$A^{BD^*}_0(m^2_K)$ &0.74 (0.65)\\
$F^{BD}_0(m^2_{K^*})$ &0.71 (0.65)&$A^{BK^*}_0(m^2_D)$ &0.40 (0.35)\\
$F^{BK}_0(m^2_{D})$ &0.42 (0.31)&$F^{BK}_1(m^2_{D^{*}})$ &0.44 (0.36)
\end{tabular}
\end{ruledtabular}
\end{table}

Starting from $A^f_{D^{(*)}P}$, FSI redistributes
these sources into the amplitudes $A_{D^{(*)}P}$, and we obtain
the corresponding rates by,
\begin{eqnarray}
\Gamma(\overline B\to PP)&=&|A_{PP}|^2  {p_{\rm cm}/(8\pi m_B^2)} ,
\nonumber\\
\Gamma(\overline B\to VP)&=&|A_{VP}|^2  {p_{\rm cm}^3/(8\pi m_V^2)} ,
\end{eqnarray}
where
\begin{equation}
2\, m_B\, p_{\rm cm}=\sqrt{[m_B^2-(m_1+m_2)^2][m_B^2-(m_1-m_2)^2]}.
\end{equation}
Note that we factor out $\varepsilon\cdot p_B$ in the definition of
the amplitude $A_{VP}$, so our expression for $\Gamma(B\to VP)$
is slightly different from that in Ref.~{\cite{Cheng:1999kd}}.


In our numerical study, we fix
$V_{ud}= 0.9749$, $V_{us}= 0.2225$, $V_{cb}= 0.04$,
and use the decay constants
$f_\pi=$ 133 MeV, $f_{K^{(*)}}=$ 158 (214) MeV, $f_{D^{(*)}}=$ 200 (230) MeV
and $f_\rho=$ 210 MeV.
Masses and lifetimes are taken from
Particle Data Group (PDG) \cite{Groom:2000in}.
We consider two form factor models:
the relativistic light-front (LF) quark model~\cite{LF} and
Neubert-Stech (NS) model~\cite{Neubert:1997uc}.
The relevant values are listed in Table~\ref{tab:formfactor}.
We use the color suppressed branching ratios of Belle~\cite{Abe:2001zi},
except for $D^{(*)0}\pi^0$ modes where
we combine with the latest CLEO numbers~\cite{Coan:2001ei}.
For other modes we use PDG values \cite{Groom:2000in}.
Since the charged $D^{(*)0}\pi^-$ mode does not rescatter to other modes, 
we normalize all modes to ${\mathcal B}(D^{(*)0}\pi^-)=
 [5.3\ (4.6)\pm0.5\ (0.4)]\times 10^{-3}$ \cite{Groom:2000in}.
We then perform a $\chi^2$ fit to the ratios of branching ratios
${\mathcal B}(D^{(*)+}\pi^-)/{\mathcal B}(D^{(*)0}\pi^-)$,
${\mathcal B}(D^{(*)0}\pi^0)/{\mathcal B}(D^{(*)0}\pi^-)$ and
${\mathcal B}(D^{(*)0}\eta)/{\mathcal B}(D^{(*)0}\pi^-)$.

The use of ratios reduces model dependence on form factors,
and is sensitive only to $a_2/a_1$.
Our numerical results for rescattering phases
in LF and NS form factor models never differ by more than  a few \%.
With $|a_2|=0.26\pm0.02$ from fit to $B\to J/\psi K$ data \cite{Cheng:1999kd}
and the range of $a_1\sim1$--1.1, $a_2\sim0.2$--0.3
from various modes \cite{Neubert:1997uc},
we shall adopt $a_2/a_1=0.25$ in subsequent discussion.
We find that a larger $a_2/a_1$ is preferred for the $D P$ modes,
but the converse is true for $D^* P$ modes.
However, for $a_2/a_1 \approx 0.25$,
the $\chi^2$s of best fits to $D P$ and $D^* P$ modes are both quite small.

\begin{table}[t!]
\caption{
\label{tab:table-phase}
The best fits in the SU(3) FSI picture.
The subscript indicates $DP$ or $D^*P$ modes.
Form factor model dependence is less than a couple of \%.}
\begin{ruledtabular}
\begin{tabular}{lccc}
      &Fit1$_{DP}$
      &Fit2$_{DP}$
      &Fit$_{D^*P}$\\
\hline
 $\chi^2$
        &0.20 \,\
        &0.27 \,\
        &0.21 \,\
        \\
 $\delta^\prime$
        &47.8$^\circ$
        &17.1$^\circ$
        &55.7$^\circ$
        \\
 $\theta$
        &24.8$^\circ$
        &$-52.7^\circ$ \,\ \
        &18.2$^\circ$
        \\
\hline
 $(1+i r_0) e^{-2i\delta_{3/2}}$
        &$0.45+0.50i$
        &$0.91+0.28i$
        &$0.32+0.47i$
        \\
 $i r_e e^{-2i\delta_{3/2}}$
        &$0.55-0.50i$
        &$0.09-0.28i$
        &$0.68-0.47i$
        \\
 $i r_a e^{-2i\delta_{3/2}}$
        &$0.14+0.04i$
        &$-0.43-0.50i$
        &$0.27-0.01i$
        \\
\hline
$|A_{1/2}/(\sqrt2 A_{3/2})|$
         &0.75 \,\
         &0.65 \,\
         &0.71 \,\
         \\
$|\delta_{1/2}-\delta_{3/2}|$
        &30.2$^\circ$
        &26.2$^\circ$
        &28.3$^\circ$
\end{tabular}
\end{ruledtabular}
\end{table}

The best fits for FSI phase differences $\delta^\prime$ and $\theta$
(or alternatively the rescattering parameters $r_i$)
are given in Table \ref{tab:table-phase}.
We do not list form factor model dependence since
it shows up often only at third decimal place.
We find two fits for the $DP$ case, but only one fit for the $D^*P$ modes.
For $DP$ modes, the set that we call ``Fit1" is similar to the $D^*P$ modes,
i.e. $\delta^\prime\sim \pm 50^\circ$, $\theta\sim \pm 20^\circ$.
We find that the quark exchange strength $|r_e|$ is
larger than the annihilation strength $|r_a|$ in this case.
As illustrated in Appendix B,
the large phase $\sim 50^\circ$ arises because of 
sizable strength of {\it both} $D^{(*)0}\pi^0$ {\it and} $D^{(*)0}\eta$.
While $\delta^\prime$ and $\theta$ effects are of similar sign for the former,
for the latter they counteract, and a large $\delta^\prime$ phase is needed.

For ``Fit2" of the $DP$ case, we have
$\delta^\prime\sim \pm 20^\circ$, $\theta\sim \mp 50^\circ$,
implying that $|r_a| > |r_e|$.
As we will see later, this fit is
ruled out by the $\overline B{}^0 \to D_s^+K^-$ bound.
We mention a curious point about our ``Pomeron" related effect,
i.e. $|1+ir_0| = 0.67$, 0.95 and 0.57, respectively,
for Fit1$_{DP}$, Fit2$_{DP}$ and Fit$_{D^*P}$.
The first and the last are remarkably consistent with
the estimates of $S_{\pi\pi\to \pi\pi} \sim 0.58$ \cite{Donoghue:1996hz},
0.68 and $S_{D\pi\to D\pi} \sim 0.76$ \cite{Suzuki:1999uc}
at $\sqrt{s} = m_B$.
In contrast, Fit2$_{DP}$, which is already ruled out by data,
is not quite consistent.

In Table \ref{tab:table-phase} we show 
$|A_{1/2}/(\sqrt2 A_{3/2})|$ and
$|\delta_{1/2}-\delta_{3/2}|$ obtained by
using the fitted strong phases.
The comparison with Eq. (\ref{eq:isospinexpt})
will be discussed in the next section.

\begin{table}[t!]
\caption{
\label{tab:table-br}
The branching ratios of various $D^{(*)}P$ modes in $10^{-4}$ units.
The second and third columns compare experiment
with factorization model.
The last two columns give the best fit results with FSI parameters
of Table \ref{tab:table-phase}.}
\begin{ruledtabular}
\begin{tabular}{lcccc}
 Mode
      &$\mathcal{B}$ ($\times 10^{-4}$)
      &fac$^{\rm LF\;(NS)}$
      &Fit1$_{DP}$
      &Fit$_{D^*P}$\\
\hline
 $\chi^2$
        &--
        &--
        &0.20
        &0.21
        \\
\hline $D^+\pi^-$
        &$30\pm4$
        &35.7 (35.5)
        &32.2
        &-- \\
$D^0\pi^0$
        & $2.9\pm 0.5$
        & 0.57 (0.58)
        & 2.93
        &-- \\
$D^0\eta$
        & $ 1.4\;^{+0.5}_{-0.4} \pm 0.3 $
        & 0.33 (0.34)
        & 1.43
        &-- \\
{$D^+_s K^-$}
        &$<2.4\ (0.7)$
        &0
        &0.03
        &-- \\
\hline
$D^{*+}\pi^-$
        &$27.6\pm2.1$
        &29.8 (29.0)
        &--
        &26.3\\
$D^{*0}\pi^0$
        & $2.5\pm0.7$
        & 0.60 (0.69)
        &--
        & 2.44
        \\
$D^{*0}\eta$
        & $ 2.0\;^{+0.9}_{-0.8} \pm 0.4$
        & 0.34 (0.39)
        & --
        & 1.83
        \\
$D^{*+}_s K^-$
        & $<1.7$
        &0
        &--
        &0.16
\end{tabular}
\end{ruledtabular}
\end{table}

We summarize the predicted rates of various modes in Table~\ref{tab:table-br}.
The branching ratios are obtained by multiplying the fitted ratios of
branching ratios by the measured central value of ${\mathcal B}(D^{(*)0}\pi^-)$.
The results for $D^{(*)+} \pi^-, D^{(*)0} \pi^0, D^{(*)0} \eta$
fit the data well, as expected from the small $\chi^2$.
It is worthy to note that rates given in Table \ref{tab:table-br}
satisfy the unitarity condition of Eq. (\ref{eq:unitary}),
i.e., the sum of branching ratios before and after FSI is equal as
shown in Eq. (\ref{eq:unitarySU3}),
up to small phase space corrections.
This clearly shows that the color suppressed modes
are fed from the $D^{(*)+}\pi^-$ mode.

At the amplitude level, one reads from Eq.~(\ref{eq:TSU3}) that 
$D^{(*)0} \pi^0$, $D^{(*)0} \eta$ and $D_s^{(*)+} K^-$ receive 
the $(r_a-r_e)/\sqrt{2}$, $(r_a+r_e)/\sqrt{6}$ and $r_a$ rescatterings, 
respectively, from $D^{(*)+} \pi^-$.
Indeed, by using phases shown in Table~\ref{tab:table-phase}
in Eq. (\ref{eq:S12SU3}),
the FSI contribution can be estimated by using
$|A_{D^{(*)0}\pi^0}/A_{D^{(*)+}\pi^-}|
\sim|{\cal S}^{1/2}_{D^{(*)+}\pi^-,D^{(*)0}\pi^0}|$
and similarly for $|A_{D^{(*)0}\eta}|$ and $|A_{D_s^{(*)+}K^-}|$.
The FSI contribution from $D^{(*)+}\pi^-$ alone provides
70\%--80\% of measured $D^{(*)0}\pi^0$, $D^{(*)0}\eta$ rates,
with the remainder coming from $a_2$ and interference terms.
For the $D_s^{(*)+}K^-$ mode, the FSI contribution from $D^{(*)+}\pi^-$
is small due to the smallness of $r_a$ but it is still
three times larger than that shown in Table~\ref{tab:table-br}.
Because of this smallness, the FSI rescattering 
from the color suppressed modes (due to the non-vanishing $a_2$) 
through $r_e$ cannot be neglected,
which reduces the rate to that shown in the Table.

We note that the factorized $a_2$ contribution to color suppressed modes
show some form factor model dependence, especially for $D^*P$ modes,
but such model dependence for fit results are rather slight.
The reduced form factor dependence is quite consistent
with FSI rescattering dominance over factorized $a_2$ amplitude,
since $D^{(*)+} \pi^-$ is the common source.

In ``Fit2" we find
$\mathcal{B}(D_s^{+} K^-)=4.65\times 10^{-4}
> \mathcal{B}(D^0\pi^0)$, $\mathcal{B}(D^0 \eta)$
due to $\vert r_a\vert > \vert r_e\vert$.
This value is above the PDG bound of
$\mathcal{B}(D_s^{+} K^-) < 2.4\times 10^{-4}$, and way above
the recent Belle bound of $0.7\times 10^{-4}$ \cite{DsK}, hence is ruled out.
The results for Fit2 are therefore not shown in Table~\ref{tab:table-br}.
On the other hand, since $|r_a|$ in the $D^{*}P$ case is not too small, 
the result of ${\mathcal B}(D_s^{*+} K^-)\sim1.6\times10^{-5}$ 
may still be of interest.

\begin{table}[b!]
\caption{
\label{tab:table-DVphase}
The best fit phase difference for $DV$ modes.
}
\begin{ruledtabular}
\begin{tabular}{lcc}
 Mode &Solution 1
      &Solution 2\\
\hline
 $\chi^2$
        &1.20 \,\
        &0.64 \,\
        \\
 $\delta^\prime$
        &18.1$^\circ$
        &--
        \\
 $\sigma$
        &--
        &34.8$^\circ$
        \\
\hline
 $(1+i r_0) e^{-2i\delta_{3/2}}$
        &$0.90+0.30i$
        &1
        \\
 $i r_e e^{-2i\delta_{3/2}}$
        &$0.10-0.30i$
        &0
        \\
 $i r_a e^{-2i\delta_{3/2}}$
        &0
        &$-0.22+0.31i$
\end{tabular}
\end{ruledtabular}
\end{table}

\subsection{FSI in $DV$ modes}

For $\overline B\to D\rho,\,D\omega$,
we have $M(\overline B\to D V)=\varepsilon\cdot p_B \, A_{DV}$
and $A^f_{D\rho}$ can be decomposed into
\begin{eqnarray}
{\it T}&=&
{G_F\over\sqrt2}V_{cb} V_{ud}^* \, a_1 \, f_\rho 2m_{\rho} F^{BD}_0(m_\rho^2),
\nonumber \\
{\it C}&=&
{G_F\over\sqrt2}V_{cb} V_{ud}^* \, a_2 \, f_{D} 2m_{\rho} A^{B\rho}_0(m_D^2),
\end{eqnarray}
while ${\it E}$ is again negligible. For $\overline B{}^0\to D^0\omega$ one 
has
\begin{equation}
A^f_{D^0\omega}=
{G_F\over\sqrt2}V_{cb} V_{ud}^* \, a_2 \, f_D 2m_{\omega} A^{B\omega}_0(m_D^2).
\end{equation}
We set $A^f_{D^{+}_s K^{*-}}=A^f_{D^0\phi}=0$ for the
factorization amplitudes of $\overline B{}^0\to D^{+}_s K^{*-},\,{D^0\phi}$ modes.
We again normalize to $D^0\rho^-$ mode since its rate is unaffected by FSI.
We take the $D^0\omega$ measurement from Ref. \cite{Abe:2001zi},
while the measurements of $D^+\rho^-$, $D^0\rho^-$ modes
and the upper limit of ${\mathcal B}(D^0\rho^0)$
are taken from PDG \cite{Groom:2000in}.

Because of the reduction to one phase difference for both
Solution 1 and Solution 2 of Eqs. (\ref{eq:Sol1}) and (\ref{eq:Sol2}),
we are able to fit with just $D^+\rho^-$ and $D^0\omega$ data.
We find $\delta^\prime$ ($\sigma$) to be
18$^\circ$ ($35^\circ$) for Solution 1 (Solution 2),
as given in Table~\ref{tab:table-DVphase}.
Since the FSI contributions to $D^0\rho^0$ and $D^0\omega$ are
mainly fed from $D^+\rho^-$
and $A^f_{D^0\rho^0}\approx -A^f_{D^0\omega}$,
Eq.~(\ref{eq:Tv}) leads to
\begin{eqnarray}
A_{D^0\rho^0(\omega)}&\approx&
{i (r^\prime_a\mp r^\prime_e)\over\sqrt2}\,
A^f_{D^+\rho^-}
+ (1+i r^\prime_0) A^f_{D^0\rho^0(\omega)}
\nonumber\\
&\approx&
{i (r^\prime_a\mp r^\prime_e)\over\sqrt2}\,
A^f_{D^+\rho^-}
\mp (1+i r^\prime_0) A^f_{D^0\omega},
\end{eqnarray}
where $r^\prime_0$, $r^\prime_a$ and $r^\prime_e$ are defined in Sec. IV.B.
For Solution~1 (Solution~2)
the dominant FSI contributions to $D^0\rho^0$ and $D^0\omega$
are the same in magnitude but opposite (same) in sign due to
$r^{(\prime)}_{e}\ (r^{(\prime)}_a) \neq 0$.
This implies different interference patterns.
For Solution 1, we have ${\mathcal B}(D^0\rho^0)\sim{\mathcal B}(D^0\omega)$.
For Solution 2, since $A^f_{D^+\rho^-}$ and $A^f_{D^0\omega}$
are real and of same sign,
\begin{eqnarray}
(1+i r^\prime_0) e^{-i \delta_{3/2}}=1,
\nonumber\\
{\rm Re}\,(i r^{\prime}_a e^{-i
\delta_{3/2}})={1\over3}[\cos\sigma-1]\leq 0,
\end{eqnarray}
the FSI contribution always interferes destructively
(contructively) with $A^f_{D^0\omega}$ ($A^f_{D^0\rho^0}$).
While ${\mathcal B}(D^0\rho^0)$ becomes larger,
one would need large annihilation contribution to account for
the observed  $D^0\omega$ data, which in turn gives rise to
${\mathcal B}(D_s^+K^{*-})$ as large as $2.7\times 10^{-4}$.
These patterns can be tested in the near future.

On the other hand, $D^0\phi$ only rescatters with $D_s^+K^{*-}$,
as can be seen from Eqs. (\ref{eq:Tv}).
It does not pick up any FSI contribution since $A^f_{D_s^+K^{*-}}=0$,
even if $A_{D_s^+K^{*-}}$ is nonvanishing as in the Solution 2 case.
Observation of $D^0\phi$ mode would imply some mechanism
at the ``source" level.
Our fitted branching ratios and predictions
for various $PV$ modes are given in Table~\ref{tab:table-DVbr}.

\begin{table}[tb]
\caption{
\label{tab:table-DVbr}
The branching ratios of various $DV$ modes.
The second and third columns compare experiment
with factorization model.
The last two columns give the best fit results with FSI parameters
of Table~\ref{tab:table-DVphase}.}
\begin{ruledtabular}
\begin{tabular}{lcccc}
 Mode & $\mathcal{B}$ ($\times 10^{-4}$)
      &fac$^{\rm LF\;(NS)}$
      &Solution 1
      &Solution 2\\
\hline
$D^+\rho^-$
        & $79\pm 18$
        & 100.7 (101.2)
        & 98.2
        & 92.7
        \\
{$D^0\omega$}
        & {$1.8\pm 0.5\;^{+0.4}_{-0.3} $ }
        & {0.67 (0.64})
        & {1.86}
        & {1.92}
        \\
{$D^0\rho^0$}
        & {$< 3.9$}
        & {0.67 (0.64})
        & {1.90}
        & {3.37}
        \\
{$D^+_s K^{*-}$}
        & $< 9.9$
        & 0
        & 0
        & {2.73}
        \\
{$D^0\phi$}
        & {--}
        & {0}
        & {0}
        & {0}
\end{tabular}
\end{ruledtabular}
\end{table}

We note that $|A_{1/2}/(\sqrt2 A_{3/2})|$=0.84,
$|\delta_{1/2}-\delta_{3/2}|=12.7^\circ$ from Solution 1
and $|A_{1/2}/(\sqrt2 A_{3/2})|$=0.81,
$|\delta_{1/2}-\delta_{3/2}|=18.5^\circ$ from Solution 2.
The phase angles are somewhat smaller than those for $D^{(*)}\pi$ modes.

\subsection{Predictions for $D^{(*)}\overline K$ and $D\overline K{}^*$ Modes}

Our FSI formulas can be applied readily to rescattering between
the Cabibbo suppressed modes, $D^{(*)+}K^-$ and $D^{(*)0}K^0$,
since they are contained in the formalism for $D^{(*)}P$.
Following similar procedure as before, we have
\begin{equation}
2\left(
\begin{array}{l}
\im\,A_{D^{(*)+}K^-}\\
\im\,A_{D^{(*)0}\overline K^{}0}
\end{array}
\right)
={\cal T}^\dagger\,
\left(
\begin{array}{l}
A_{D^{(*)+}K^-}\\
A_{D^{(*)0}\overline K{}^0}
\end{array}
\right),
\label{eq:2ImA=TADK}
\end{equation}
where
\begin{equation}
{\cal T}=\left(
\begin{array}{cc}
r_0 &r_e\\
r_e &r_0
\end{array}
\right),
\label{eq:TDK}
\end{equation}
and ``annihilation" is clearly impossible as can be seen from the pictorial
approach.
By using Eq.~(\ref{eq:r1r2SU3}), 
we note that ${\cal S}={\bf 1}+i{\cal T}$ is automatically unitary. 
Therefore, we obtain
\begin{equation}
\left(
\begin{array}{c}
A_{D^{(*)+}K^-}\\
A_{D^{(*)0}\overline K{}^0}
\end{array}
\right)
={\cal S}^{1/2}\,
\left(
\begin{array}{c}
A^f_{D^{(*)+}K^-}\\
A^f_{D^{(*)0}\overline K^{}0}
\end{array}
\right),
\label{eq:FSIDK}
\end{equation}
where, 
\begin{equation}
{\cal S}^{1/2}={e^{i\delta_{3/2}}\over2}
\left(
\begin{array}{cc}
1+e^{i\delta^\prime} &1-e^{i\delta^\prime}\\
1-e^{i\delta^\prime} &1+e^{i\delta^\prime}
\end{array}
\right),
\end{equation}
which is consistance with those obtained in Appendix A,
and the factorization amplitudes are
\begin{eqnarray}
A^f_{D^+K^-}&=&{G_F\over\sqrt2} V_{cb} V_{us}^* \, a_1 \, (m_B^2-m_D^2) f_K
           F_0^{BD}(m^2_K),
\nonumber\\
A^f_{D^0\overline K{}^0}
	    &=&{G_F\over\sqrt2} V_{cb} V_{us}^* \, a_2 \, (m_B^2-m_{K^0}^2) f_D
           F_0^{BK}(m^2_{D^0}),
\nonumber \\
A^f_{D^{*+}K^-}&=&{G_F\over\sqrt2} V_{cb} V_{us}^* \, a_1 \, 2 m_{D^*} f_K
           A_0^{BD}(m^2_K),
\nonumber\\
A^f_{D^{*0}\overline K{}^0}&=&{G_F\over\sqrt2} V_{cb} V_{us}^* \, a_2 \, 2 m_{D^*} f_{D^*}
           F_1^{BK}(m^2_{D^{*0}}),
\end{eqnarray}
where form factors are found in Table~\ref{tab:formfactor}.
One again has a triangle relation
$A^f_{D^{(*)0}K^-}=A^f_{D^{(*)+}K^-}+A^f_{D^{(*)0}\overline K{}^0}$.

\begin{table}[t!]
\caption{
\label{tab:table-DK}
The predicted branching ratios of $D^{(*)}\overline K$ and $D \overline K{}^*$
modes in $10^{-4}$ units.
The second and third columns compare experiment
with factorization model.
The last three columns give the best fit results with FSI parameters
of Table \ref{tab:table-phase} and Table \ref{tab:table-DVphase}.}
\begin{ruledtabular}
\begin{tabular}{lccccc}
 Mode
      &$\mathcal{B}$ ($\times 10^{-4}$)
      &fac$^{\rm LF
		   }$
      &Fit1$_{DP}$
      &Fit$_{D^*P}$
      &Solution~1\\
\hline
$D^0 K^-$
        &$4.2\pm0.7$
        &4.12 
        &--
        &--
        &-- \\
$D^+K^-$
        & $2.0\pm 0.6$
        & 2.61 
        & 2.20
        & --
        & -- \\
$D^0 \overline K{}^0$
        &--
        & 0.12 
        & 0.53
        & --
        &--\\
\hline
$D^{*0} K^-$
        &$3.6\pm1.0$
        &3.40 
        & --
        & --
        & --\\
$D^{*+}K^-$
        & $2.0\pm0.5$
        & 2.15 
        &--
        & 1.71
        & --
        \\
$D^{*0}\overline K{}^0$
        & --
        & 0.10 
        & --
        & 0.55
        & --\\
\hline
$D^0 K^{*-}$
        &--
        &7.0 
        & --
        & --
        & --
\\
$D^+K^{*-}$
        &--
        & 5.11 
        &--
        &--
        & 4.99 
\\
$D^0\overline K{}^{*0}$
        & --
        & 0.09 
        & --
        & --
        & 0.21 
\end{tabular}
\end{ruledtabular}
\end{table}

The $\delta^\prime$ phase has already been fitted, and the predicted branching ratios 
for  $D\overline K$ and $D^*\overline K$ modes are given in Table~\ref{tab:table-DK}.
The second column is obtained by multiplying Belle measurements of
${\mathcal B}(D^{(*)+,0}K^-)/{\mathcal B}(D^{(*)+,0}\pi^-)$~\cite{Abe:2001wa}
by PDG values of ${\mathcal B}(D^{(*)+,0}\pi^-)$~\cite{Groom:2000in}.

It is interesting to note that $D^{(*)}\overline K$ modes do not receive
the annihilation type FSI, $r_a$,
and the $\overline K{}^0$ wave function does not have the
$1/\sqrt2$ factor as compared to $\pi^0$.
For ``Fit1$_{\rm DP}$" and ``Fit$_{\rm D^* P}$,
$r_a$ is subdominant while $|r_e|$ is close to each other, hence
we find that $\mathcal{B}(D^{0}\overline K{}^0)/\mathcal{B}(D^{+}K^-) \approx
2\mathcal{B}(D^{0}\pi^0)/\mathcal{B}(D^{+}\pi^-)\approx 2\times 1/10$
and $\mathcal{B}(D^{*0}\overline K{}^0)\approx \mathcal{B}(D^{0}K^0)$.
As noted, ``Fit2$_{\rm DP}$" is ruled out already by $D_s^+K^-$ bound.

In a very similar fashion, we predict the rescattering of
$D^+K^{*-}$ into $D^0\overline K{}^{*0}$ final state, which is given again in 
Table~\ref{tab:table-DK} for Solution 1 of $DV$ case, where one again has 
$\mathcal{B}(D^{0}\overline K{}^{*0})/\mathcal{B}(D^{+}K^{*-}) \approx
2\mathcal{B}(D^{0}\rho^0)/\mathcal{B}(D^{+}\rho^-)\approx 2\times 1/50$.
For Solution 2, $r_e = 0$ and the result is the same as
the second column for factorization.

The $D\overline K{}^{*}$ modes have yet to be observed.
The factorization predictions for $D^0K^{*-}$ and $D^+K^{*-}$
are about twice as large as $D^{*0}K^-$ and $D^{*+}K^-$,
but the predicted branching ratio for $\overline B \to D^0\overline K{}^{*0}$
is less than half of $\overline B \to D^{*0}\overline K{}^0$ in Solution 1.
This is because $\vert r_e\vert$ (or the rescattering phase $\delta^\prime$)
for $DV$ modes are much smaller than for $D^*P$ modes.
We note that in Solution 2 one would predict
$\overline B \to D^0\overline K{}^{*0}$ to occur at half the rate of 
the Solution 1 case,
i.e. just the factorization $a_2$ prediction.

Our formalism therefore predicts a relatively sizable
$\mathcal{B}(D^{(*)0}\overline K{}^0)$ at $\sim 0.5\times 10^{-4}$,
and expect $\mathcal{B}(D^0\overline K{}^{*0})\sim 0.2\times 10^{-4}$
($0.1\times 10^{-4}$ if Solution 2 is confirmed).
We encourage Belle and BaBar to search for these modes.

\section{Discussion and Conclusion}

We have stressed that the isospin relation of Eq.~(\ref{eq:triangle}),
which follows from Eq.~(\ref{eq:isospin}) 
(given more explicitly in Eqs.~(\ref{eq:isospina})--(\ref{eq:isospinc})),
holds whether one has FSI or not.
This was used to extract $A_{1/2}/A_{3/2}$ directly from $D^{(*)}\pi$ data, 
as given in Eq.~(\ref{eq:isospinexpt}), which we reproduce here:
\begin{eqnarray}
{1\over\sqrt2} \left\vert {A_{1/2}\over A_{3/2}}\right\vert^{D^{(*)}\pi{\rm\ only}}
&=&0.71\pm0.11\ (0.75\pm0.08),
\nonumber \\
|\delta|^{D^{(*)}\pi{\rm\ only}}
&=&29^\circ\pm6^\circ\ (30^\circ\pm7^\circ),
\label{eq:Eq16}
\end{eqnarray}
where it is made clear that they are extracted from $D^{(*)}\pi$ data alone.
On the other hand, we have given in Table \ref{tab:table-phase}
the values for $|A_{1/2}/(\sqrt2 A_{3/2})|$ and $|\delta_{1/2}-\delta_{3/2}|$
as obtained by using the fitted strong phases that 
takes into account $D^0\eta^{(*)0}$ data,
i.e. by using the $A_{D^{(*)0}\pi^-}$, $A_{D^{(*)+}\pi^-}$ 
and $A_{D^{(*)0}\pi^0}$ amplitudes of Eq.~(\ref{eq:FSISU3}).
They turn out to be not so different from Eq.~(\ref{eq:Eq16}). 
Let us understand why.

We note that $|A_{1/2}/(\sqrt2 A_{3/2})| = 1 + {\cal O}(\Lambda/m_Q) \longrightarrow 1$
in  the heavy $b$ (and $c$) quark limit \cite{Beneke:2000ry,Neubert:2001sj},
although $m_c$ may not be heavy enough.
The strong phase $|\delta_{1/2}-\delta_{3/2}|\longrightarrow 0$ 
in the absence of short and long distance rescattering.
It is intructive to consider first 
the elastic $D^{(*)}\pi\to D^{(*)}\pi$ rescattering case. 
Noting that elastic rescattering 
does not change $|A_I|$ from its factorization value, 
for $D^{(*)}P$ modes we have, roughly speaking,
\begin{equation}
{|A_{1/2}|\over\sqrt2 |A_{3/2}|}=
{|A^f_{1/2}|\over\sqrt2 |A^f_{3/2}|}={|2 T-C|\over 2|T+C|}
\sim{|2-\frac{a_2}{a_1}|\over2|1+\frac{a_2}{a_1}|},
\label{eq:a2}
\end{equation}
which deviates from 1 due to a nonzero $a_2$,
which is a nonfactorizable effect.
It was a happy coincidence, before the measurement of color-suppressed modes,
that taking $a_2/a_1=0.25$ and {\it real} could account for \cite{Neubert:1997uc}
both $B^- \to D^0\pi^-$ and $\overline B\to J/\psi \overline K{}^{(*)}$ rates.
It should be stressed that the sizable value of $|a_2/a_1|$
can be viewed as determined this way from data
that gives $|A_{1/2}/(\sqrt2 A_{3/2})| \simeq 0.7$.

The impact of the new experimental measurement of $D^0\pi^0$
is that one is now able to determine the strong phase difference,
$|\delta_{1/2}-\delta_{3/2}|\sim 30^\circ$, which is not quite small.
With this one has two ways to proceed.

As mentioned in the Introduction,
Refs.~\cite{Neubert:2001sj,Cheng:2001sc}
continue to employ factorization formulas to $D^{(*)0}\pi^0$ 
hence make $\vert a_2\vert$ larger by roughly {\it a factor of two}.
To maintain $D^{(*)0}\pi^-$ {\it and} generate $\delta_{1/2}-\delta_{3/2}$,
one capitalizes on Eq.~(\ref{eq:a2}) 
by dropping the absolute value condition.
That is, one resorts to a complex $a_2/a_1$ itself.
In this way one finds $\vert a_2\vert \sim 0.4$--0.5 and $\arg a_2 \sim 60^\circ$
could account for $\overline B \to D\pi$ data, at the cost that 
$\vert a_2\vert$ is twice as large as from $J/\psi \overline K{}^{(*)}$ modes.
Ref. \cite{Neubert:2001sj} argues further that,
while factorization no longer holds, the trend of larger and complex $a_2$ is 
expected from QCD factorization \cite{Beneke:2000ry}.

Our critique is the following.
First, 
this ``$a_2$" approach is process dependent, and predictiveness is lost.
Although $\vert a_2\vert \sim 0.4$--0.5 could account for the strength of
observed $D^{(*)0}\eta$ and $D^{(*)0}\omega$ modes,
it seems coincidental, with $\vert a_2\vert$ varying by $\sim 20$--30\% 
among these and $D^{(*)0}\pi^0$ modes, while we know that 
$\vert a_2\vert \sim 0.2$--0.3 for $J/\psi \overline K{}^{(*)}$ modes.
Second, 
it is the need to maintain $D^{(*)0}\pi^-$ rate
that a sizable phase to $a_2$ is invoked,
although previously a smaller and real $a_2$ gave a pleasant, consistent picture.
However, from Eq. (\ref{eq:a2}) we know that $a_2$ reduces $|A_{1/2}/(\sqrt2 A_{3/2})|$ 
hence it represents {\it inelastic} effects \cite{Suzuki:2000ej}.
This is the reason why it is not quite calculable.
In this sense, however, $\arg a_2 \sim 60^\circ$
is not reasonable since one expects strong cancellations
among numerous inelastic channels \cite{Suzuki:1999uc,Gerard:1991ni}.
A statistical model suggests the typical phase to be
$\sim 20^\circ$ \cite{Suzuki:1999uc}.
Third, 
we stress that generating Eq.~(\ref{eq:Eq16}) by 
the phase and strength of $a_2$ holds only when 
one drops the absolute value requirement from Eq. (\ref{eq:a2}),
i.e. ignoring elastic FSI, as is common practice in 
QCD factorization \cite{Beneke:2000ry}.
Such FSI effects are ${\cal O}(\alpha_s(m_b))$ suppressed, or $1/m_Q$ suppressed.
For the former, clearly $\alpha_s(m_b) \sim 0.2 \sim 10^\circ$ in radians.
For the latter, one has the real problem that $m_c$ may {\it not} be heavy enough.

The approach advocated in this paper is via quasi-elastic FSI rescattering.
Let's make a point by point comparison with the ``process dependent $a_2$" approach.
First, 
process independence is not so easily conceded.
In particular, we maintain $a_2/a_1 \cong 0.25$ and real.
Thus, the proximity of Eq. (\ref{eq:a2}) with Eq.~(\ref{eq:Eq16})
reflects a phenomenological tuning done several years ago.
The philosophy is that, while we agree that $a_2$ should be process dependent
and in principle complex, we take the above ``tuned" value as no mere accident. 
That is, Nature could have revealed to us long ago that 
$a_2$ is strongly process dependent.
Since keeping both $a_2$ as parameter {\it and} allowing for FSI phases
cannot lead one afar, we opt to keep $a_2/a_1$ fixed as done in
\cite{Neubert:1997uc}.
The new experimental measurement of $D^{(*)0}\eta$ and $D^{(*)0}\omega$ modes,
rather than giving ``process dependence" irritation,
can be incorporated nicely by enlarging the scope of FSI
from elastic SU(2) to quasi-elastic SU(3) symmetry.
This stretching of ``elasticity",
together with maintaining process independence to good degree, 
make our approach suitably predictive.
Second,
by leaving $a_2/a_1$ as done before in \cite{Neubert:1997uc},
one enjoys the success with $D^0\pi^-/D^+\pi^-$ and $J/\psi \overline K$.
The strength of $a_2$ is smaller, hence the acuteness of inelasticity
is not as severe as the ``$a_2$" approach.
A strong phase of order $50^\circ$ does emerge,
but this is interpretted as due to having
$D^{(*)0}\pi^0$ and $D^{(*)0}\eta$ {\it both sizable} (see Appendix B),
and has completely different origins
from the need for $\arg a_2 \sim 60^\circ$ in the large ``$a_2$" approach.
Finally, in comparison with the strong assumption of 
removing the absolute value condition from Eq. (\ref{eq:a2}),
we took advantage of FSI approach to expand a previously commonly known 
``folklore" on elastic FSI rescattering.

We can now comment on comparison of Eq. (\ref{eq:Eq16}) with values in Table II.
With $a_2/a_1 = 0.25$, we find
$|A_{1/2}/(\sqrt2 A_{3/2})| =|A^f_{1/2}/(\sqrt2 A^f_{3/2})| =0.77\ (0.75)$
for LF and $0.77\ (0.73)$ for NS form factors. 
Assuming elastic rescattering and using the formulas of Sec. III,
the strength of amplitudes cannot change,
and we find $|\delta_{1/2}-\delta_{3/2}|=30.6^\circ\ (29.9^\circ)$ for LF
and $30.6^\circ (29.2^\circ)$ for NS form factors.
Note that the form factor dependence is very weak.
With rescattering among SU(3) multiplets, $A^f_{1/2}$ can now feed
other color suppressed modes via rescattering (Eq.~(\ref{eq:S12block})).
The $D^{(*)0}\pi^-$ mode still can not rescatter to other modes, 
so $|A_{3/2}|=A^f_{3/2}=\vert A_{D^0\pi^-}\vert/\sqrt3$. 
Thus, from unitarity we expect $|A_{1/2}/A_{3/2}|<|A^f_{1/2}/A^f_{3/2}|$
in the presence of quasi-elastic FSI.
We see from Table \ref{tab:table-phase} that $|A_{1/2}/A_{3/2}|$ is 
reduced from the factorization results by 3\%, 16\% respectively, 
for Fit1$_{DP}$, Fit2$_{DP}$, and by 5\% for Fit$_{D^*P}$.
Except the second case, which is ruled out by $\overline B{}^0 \to D_s^+K^-$ data,
the deviation from elastic FSI is mild.
This reflects the fact that the color-suppressed $D^{(*)0}\eta$ rate
is still small compared to $D^{(*)+}\pi^-$.
The strong phases $|\delta_{1/2}-\delta_{3/2}|$ in Table~\ref{tab:table-phase}
agree rather well with the directly extracted ones (Eq. (\ref{eq:Eq16})) 
as well as the elastic ones,
and the validity of Eq.~(\ref{eq:estimate}) as good estimates is born out.

A principle motivation and interest in understanding 
color-suppressed $\overline B \to D^{(*)0}h^{(*)0}$ modes is its 
possible implication for $\overline B\to \overline K\pi$ and $\pi\pi$ final states.
These modes have been one of the focal points in $B$ physics in recent years,
because it provides rich probes of CP violation~\cite{He:1999mn}
and possibilities~\cite{He:1998ej} for new physics.
We note that the effect of $a_2$ is rather subdued in these processes,
but our picture of rescattering may still be realized,
hence these processes provide more fertile testing ground for FSI.
Effects of FSI rescattering on $K\pi$, $\pi\pi$ final states have been discussed in the 
literature \cite{Suzuki:1999uc,Gerard:1991ni,Donoghue:1996hz,Falk:1998wc}.
In particular, it has been stressed \cite{Hou:1999st} that
large rescattering phases in $\overline K\pi \to \overline K\pi$ and
$\pi\pi\to \pi\pi$ could have dramatic impact on such charmless final states.
If phases analogous to $\delta^\prime \sim 50^\circ$ can be realized,
the $\overline K{}^0\pi^0$ and $\pi^0\pi^0$ modes could get enhanced 
while $\pi^+\pi^-$ mode suppressed.
Direct CP rate asymmetries could soon be observed in $\overline K\pi$ modes,
in particular in ``pure penguin" $\overline K{}^0\pi^-$ mode,
while for $\pi^0\pi^0$ and $\pi^+\pi^-$ modes they could even reach 50--60\%.
The formalism is a straightforward extension from the one presented here.
Rather than ${\rm D}\Pi \to {\rm D}\Pi$ rescattering,
one now needs to study $\Pi\Pi \to \Pi\Pi$ rescattering.
It is interesting to note that the factorization ``sources" 
for all four $\overline K\pi$ modes are sizable,
unlike our present case where the $D^+\pi^-$ mode is the singly large source.
Thus, the cross-feed between channels would be different.
Furthermore, the CP violating rate asymmetries provide additional leverage to check
for the presence of FSI phases.
In this sense the $K\pi$, $\pi\pi$ ($PP$ in broader sense) system
is richer than our present $D^{(*)}P$ case.
It is interesting that the physical picture of $r_i$ is still applicable 
with an additional annihilation rescattering term, 
due to possible final states consist of $P\bar P$.
Our study is underway and would be reported elsewhere.

Finally, we mention a curiosity. Fit2$_{DP}$ contains large ``annihilation" rescattering, 
which runs against $\overline B{}^0\to D_s^+K^-$ data hence is ruled out.
Fit1$_{DP}$ and Fit$_{D^*P}$, as well as Solution 1 of $DV$ case,
all had exchange rescattering far dominant over annihilation.
Thus, Solution 2 of $DV$ case is the only one where the latter is sizable and dominant.
It therefore has the distinct feature that
$D^0\rho^0$ is almost twice as large as $D^0\omega$,
with $D_s^+K^-$ not much smaller.
However, if this were realized,
then one would expect $\overline B{}^0 \to D^0\overline K{}^{*0}$ to be
at factorization rate and much weaker than
$D^0\overline K{}^{0}$ and $D^{*0}\overline K{}^{0}$,
which would be rather peculiar.
We would therefore not be surprised if Solution 2 gets ruled out soon,
and one might then conclude that rescattering is largely in terms of
the classic ``charge exchange" type.
This may also explain why $K\overline K$ modes are so far unseen.
In this vein we wish to remark also that we 
have not exhausted the predictiveness of our approach.
For instance, one could generate 
``wrong charge" $\overline B \to \overline D^0 \overline K$ decays via
$D_s^-\pi^0\to \overline D^0 K^-$ and 
$D_s^-\pi^+\to \overline D^0 \overline K{}^0$ rescattering
from $V_{ub}$ suppressed $\overline B \to D_s^-\pi^0$, $D_s^-\pi^+$ decays.
The rescattering matrix can be adapted from results presented here,
but the latter decays have yet to be observed.

In conclusion,
we advocate in this work the possibility that
the recently observed host of $\overline B\to D^0h^0$ modes
may be hinting at final state rescattering.
In contrast to a suggestion of a larger and complex $a_2$,
we extend the elastic $D^{(*)}\pi\to D^{(*)}\pi$ FSI picture
to quasi-elastic $D^{(*)}P \to D^{(*)}P$ and $DV \to DV$ rescattering,
where $P$ is the pseudoscalar SU(3) octet, and $V$ is the vector U(3) nonet.
In this way we are able to accommodate $D^{(*)0}\eta$, $D^{0}\omega$ modes
in a unified setting.
For $D^{(*)}P$ modes, we find that data give rise to two rescattering phases
$\delta^\prime \sim 50^\circ$ and $\theta \sim 20^\circ$,
where the need for a large phase comes about because of
the strength of {\it both} the $D^{(*)0}\pi^0$ and $D^{(*)0}\eta$ modes.
For $DV$ modes, nonet symmetry reduces the number of physical phases to one,
of order 20--30$^\circ$.
The emerging pattern is that of ``charge exchange" rescattering,
rather than ``quark annihilation".
We predict rather small $\overline B{}^0 \to D_s^{(*)+}K^-$, $D_s^+K^{(*)-}$ 
and $D^0\phi$ rates, and $D^0\rho^0 \simeq D^0\omega$,
although in one solution one could have
$D^0\rho^0 \simeq 2\times D^0\omega$ and
$\overline B{}^0 \to D_s^+K^{(*)-}\sim 3\times 10^{-4}$,
which can be easily checked.
We expect $\overline B{}^0 \to D^{(*)0}\overline K{}^0$ 
and $D^0\overline K{}^{*0}$ to be at $0.55\times 10^{-4}$
and $0.2\times 10^{-4}$, respectively, which is sizable.
While these predictions can be tested experimentally,
the $K\pi$, $\pi\pi$ charmless final states are even more promising,
if FSI phases are as large as $50^\circ$,
because one then expects rather sizable direct CP asymmetries
with a distinct pattern.

\begin{acknowledgments}
This work is supported in part by
the National Science Council of R.O.C.
under Grants NSC-90-2112-M-002-022, NSC-90-2811-M-002-038
and NSC-90-2112-M-033-004,
the MOE CosPA Project,
and the BCP Topical Program of NCTS.
\end{acknowledgments}

\begin{widetext}

\appendix

\section{SU(3) Decomposition of the Rescattering matrix}

It is well known that $D^{0,+},\,D^+_s$ and $\pi,\,K,\,\eta$ transform respectively
as $\overline {\bold 3}$ and $\bold 8$ under SU(3), 
\begin{equation}
{\overline D}(\overline 3)=
\left(
\begin{array}{ccc}
D^0 &D^+ &D^+_s
\end{array}
\right),
\qquad
{\Pi(8)}=
\left(
\begin{array}{ccc}
{{\pi^0}\over\sqrt2}+{{\eta_8}\over\sqrt6}&{\pi^+} &{K^+}\\
{\pi^-}&-{{\pi^0}\over\sqrt2}+{{\eta_8}\over\sqrt6}&{K^0}\\
{K^-}  &{\overline K{}^0}&-\sqrt{2\over3}{\eta_8}
\end{array}
\right).
\end{equation}
The $\overline D(\overline 3)\otimes\Pi(8)$ can be reduced into a 
$\overline{\bold 3}$, a $\bold 6$ and a $\overline{\bold 15}$, 
i.e. (see for example, \cite{georgi}) 
\begin{eqnarray}
T(\overline 3)_j\equiv \overline D_l\Pi^l_j,
\quad
T(6)^{li}\equiv\epsilon^{lmn} \overline D_m \Pi^i_n+
                 \epsilon^{imn} \overline D_m \Pi^l_n,
\nonumber\\
T(\overline {15})^i_{jk}\equiv \overline D_k \Pi^i_j+\overline D_j \Pi^i_k
         -{1\over4} \delta^i_k \overline D_l \Pi^l_j
         -{1\over4} \delta^i_j \overline D_l \Pi^l_k.
\label{eq:TTT}
\end{eqnarray}
The SU(3) symmetry of strong interaction enforces the twenty-four by twenty-four
scattering matrix having the following form
\begin{equation}
{\cal S}^{1/2}=e^{i\delta_{\overline{15}}}\sum_{a=1}^{15} |T(\overline {15});a\rangle\langle T(\overline {15});a|
+e^{i\delta_{{6}}}\sum_{b=1}^{6} |T(6);b\rangle\langle T(6);b|
+e^{i\delta_{\overline{3}}}\sum_{c=1}^{3} |T(\overline 3);c\rangle
\langle T(\overline3);c|;
\label{eq:diagonal}
\end{equation}
where $|T(\overline {15});a\rangle$, $|T(6);b\rangle$ and $|T(\overline 3);c\rangle$
are orthonormal SU(3) basis for the irreducible representations
shown in the above equation.

Although the twenty-four by twenty-four rescattering matrix is diagonal in these basis,
we may not need all of them in a realistic situiation.
For a final state with given strangness and isospin (or electric charge)
it can only rescatter to other final states having same quantum numbers.
For later purpose, we give the explicit forms of these basis.
By proper linear combining states within the same multiple, as shown in 
Eq. (\ref{eq:TTT}),
$|T(\overline {15});a\rangle$ can be classified according to strangthness
($S$) and isospin ($I$)
\begin{eqnarray}
(S=1,\,I=1)&:&\frac{|D^+K^+\rangle+|D_s^+\pi^+\rangle}{\sqrt2},
\frac{|D^+K^0\rangle-|D^0K^+\rangle-\sqrt2 |D^+_s\pi^0\rangle}{2},
\quad
\frac{|D^0K^0\rangle+|D_s^+\pi^-\rangle}{\sqrt2};
\\
(S=1,\,I={1\over2})&:&|D_s^+K^+\rangle,\quad|D_s^+K^0\rangle;
\\
(S=1,\,I=0)&:&
\frac{|D^+K^0\rangle+|D^0K^+\rangle+\sqrt6 |D_s^+\eta_8\rangle}{2\sqrt2}.
\\
(S=0,\,I={3\over2})&:&|D^+\pi^+\rangle,\quad
{1\over3}|D^0\pi^+\rangle+\sqrt{2\over3}|D^+\pi^0\rangle,\quad
{1\over\sqrt3}|D^+\pi^-\rangle-\sqrt{2\over3}|D^0\pi^0\rangle,\quad
|D^0\pi^-\rangle;
\\
(S=0,\,I={1\over2})&:&\frac{2|D^0\pi^+\rangle-\sqrt2 |D^+\pi^0\rangle+3\sqrt6|D^+\eta_8\rangle
     -6|D^+_s\overline K{}^0\rangle}{4\sqrt6},
\\
&&\frac{2|D^+\pi^-\rangle+\sqrt2 |D^0\pi^0\rangle+3\sqrt6|D^0\eta_8\rangle
     -6|D^+_sK^-\rangle}{4\sqrt6};
\\
(S=-1,\,I=1)&:&|D^+\overline K{}^0\rangle,
\quad
\frac{|D^+K^-\rangle+|D^0\overline K{}^0\rangle}{\sqrt2},
\quad
|D^0K^-\rangle.
\end{eqnarray}

Similarly $|T(6),b\rangle$ are
\begin{eqnarray}
(S=1,\,I=1)&:&\frac{|D^+K^+\rangle-|D_s^+\pi^+\rangle}{\sqrt2},\quad
   \frac{|D^+K^0\rangle-|D^0K^+\rangle+\sqrt2|D^+_s\pi^0\rangle}{2},\quad
\frac{|D^0K^0\rangle-|D_s^+\pi^-\rangle}{\sqrt2},
\\
(S=0,\,I={1\over2})&:&
\frac{2|D^0\pi^+\rangle-\sqrt2 |D^+\pi^0\rangle-\sqrt6|D^+\eta_8\rangle
     -2|D^+_s\overline K{}^0\rangle}{4},
\nonumber\\
&&\frac{2|D^+\pi^-\rangle+\sqrt2 |D^0\pi^0\rangle-\sqrt6|D^0\eta_8\rangle
     -2|D^+_sK^-\rangle}{4},
\\
(S=-1,\,I=0)&:&
\frac{|D^+K^-\rangle-|D^0\overline K{}^0\rangle}{\sqrt2}.
\end{eqnarray}

The $|T(\overline 3),c\rangle$ are 
\begin{eqnarray}
(S=0,\,I=1/2)&:&
\frac{6|D^0\pi^+\rangle-3\sqrt2 |D^+\pi^0\rangle+\sqrt6|D^+\eta_8\rangle
     +6|D^+_s\overline K{}^0\rangle}{4\sqrt6},
\nonumber\\
&&\frac{6|D^+\pi^-\rangle+3\sqrt2 |D^0\pi^0\rangle+\sqrt6|D^0\eta_8\rangle
     +6|D^+_sK^-\rangle}{4\sqrt6},
\\
(S=1,\,I=0)&:& 
\frac{3|D^+K^0\rangle+3|D^0K^+\rangle-\sqrt6|D^+_s\eta_8\rangle}{2\sqrt6}.
\end{eqnarray}

With these basis, it is then straightfarward to obtain ${\cal S}^{1/2}$
by using Eq. (\ref{eq:diagonal}).
As noted before we only need final states with same quantum numbers for 
rescattering.
For example, $I=3/2$ states can only appear in $\overline{15}$, hence
we identify $\delta_{3/2}$ as $\delta_{\overline{15}}$.
For $S=-1$ and $Q=0$, we only have 
$D^0 \overline K{}^0$ and $D^+K^-$ for rescattering.
By using two neutral and $S=-1$ states in $\overline{15}$ and $6$,
respectively, we immediately obtain ${\cal S}^{1/2}$ in Eq. (\ref{eq:FSIDK}). 
Similarly, by using $S=Q=0$, $I=3/2,\,1/2$ states in $\overline{15}$
and $I=1/2$ states in $6$ and $\overline3$,
we obtain ${\cal S}^{1/2}$ in Eq. (\ref{eq:S12SU3}) readily.  

We now turn to the $DV$ case. 
The SU(3) decomposition of $DV$ final states can be obtained by
replacing $\pi$, $K$ and $\eta_8$ in the $DP$ case  
by $\rho$, $K^*$ and $\omega_8$, respectively.
However, there is another $\overline {\bold 3}$ 
as $|T({\overline 3{}^\prime});i\rangle=|\overline D(\overline 3)_i\omega_1\rangle$.
This $\overline {\bold 3}{}^\prime$ can mix with the previous $\overline {\bold 3}$ 
with a two by two symmetric (due to time reversal invarint) unitary matrix 
${\cal U}^{1/2}$. We have

\begin{equation}
{\cal S}^{1/2}_V=e^{i\delta_{\overline{15}}}\sum_{a=1}^{15} |T(\overline {15});a\rangle\langle T(\overline {15});a|
+e^{i\delta_{{6}}}\sum_{b=1}^{6} |T(6);b\rangle\langle T(6);b|
+\sum_{m,n=\overline3,\overline3{}^\prime}\sum_{c=1}^{3} 
|T(m);c\rangle{\cal U}^{1/2}_{mn}
\langle T(n);c|,
\label{eq:diagonalDV}
\end{equation}
for $DV$ rescatering matrix.
Note that the symmetric mixing matrix ${\cal U}^{1/2}$ can be parametrized 
by two phases and one mixing angle.
We need four paramters, including three phase differences and one mixing angle,
to discribe the $DV$ FSI case.
On the other hand by nonet symmetry we reduce them to only one parameter  
(in addition to an overall phase) but with two distinct solutions as discussed 
in Section IV.

\section{geometric representation}

We give the geometric (triangular) representation of
our results in this Appendix.
For simplicity of presentation,
we consider the leading FSI contribution and drop the $a_2$ contribution.
The triangular relation for the $D\pi$ system,
Eqs. (\ref{eq:isospina}) and (\ref{eq:isospinb}),
and the FSI formula for $D^0\eta$, $D_s^+K^-$,
Eqs. (\ref{eq:FSIblock}) and (\ref{eq:S12block}),
will then be reduced to
\begin{eqnarray}
\sqrt3 A_{D^+\pi^-}=A_{3/2}+\sqrt2 A_{1/2},
&\qquad&
\sqrt2 A_{1/2}={1\over16} 
                 (1+6 e^{i\delta^\prime}+9 e^{i\theta})
                 (\sqrt2 e^{i\delta_{3/2}}A^f_{1/2}),
\nonumber\\
\sqrt6 A_{D^0\pi^0}=-2A_{3/2}+\sqrt2 A_{1/2},
&\qquad&
\sqrt2 A_{D^0\eta}={1\over16} 
                 (3-6 e^{i\delta^\prime}+3 e^{i\theta})
                 (\sqrt2 e^{i\delta_{3/2}}A^f_{1/2}),
\nonumber\\
\sqrt3 A_{D^0\pi^-}=3 A_{3/2},
&\qquad&
{A_{D_s^+K^-}\over\sqrt3}={1\over16} 
                 (-1-2 e^{i\delta^\prime}+3 e^{i\theta})
                 (\sqrt2 e^{i\delta_{3/2}}A^f_{1/2}), 
\nonumber\\
A_{3/2}=e^{i\delta_{3/2}} A_{3/2}^f,
&\qquad&
\sqrt2 A^f_{1/2}=2A_{3/2}^f,
\end{eqnarray}
where the last equation follows from
$\sqrt6 A^f_{D^0\pi^0} = - 2 A^f_{3/2} + \sqrt2 A^f_{1/2}=0$ when $a_2=0$.
Eq. (\ref{eq:triangle}) still holds, and $|A_{3/2}|=|A_{3/2}^f|$,
but we now have $|A_{1/2}|\leq |A^f_{1/2}|$ due to quasi-elastic rescattering.

\begin{figure}[t!]
\centerline{\DESepsf(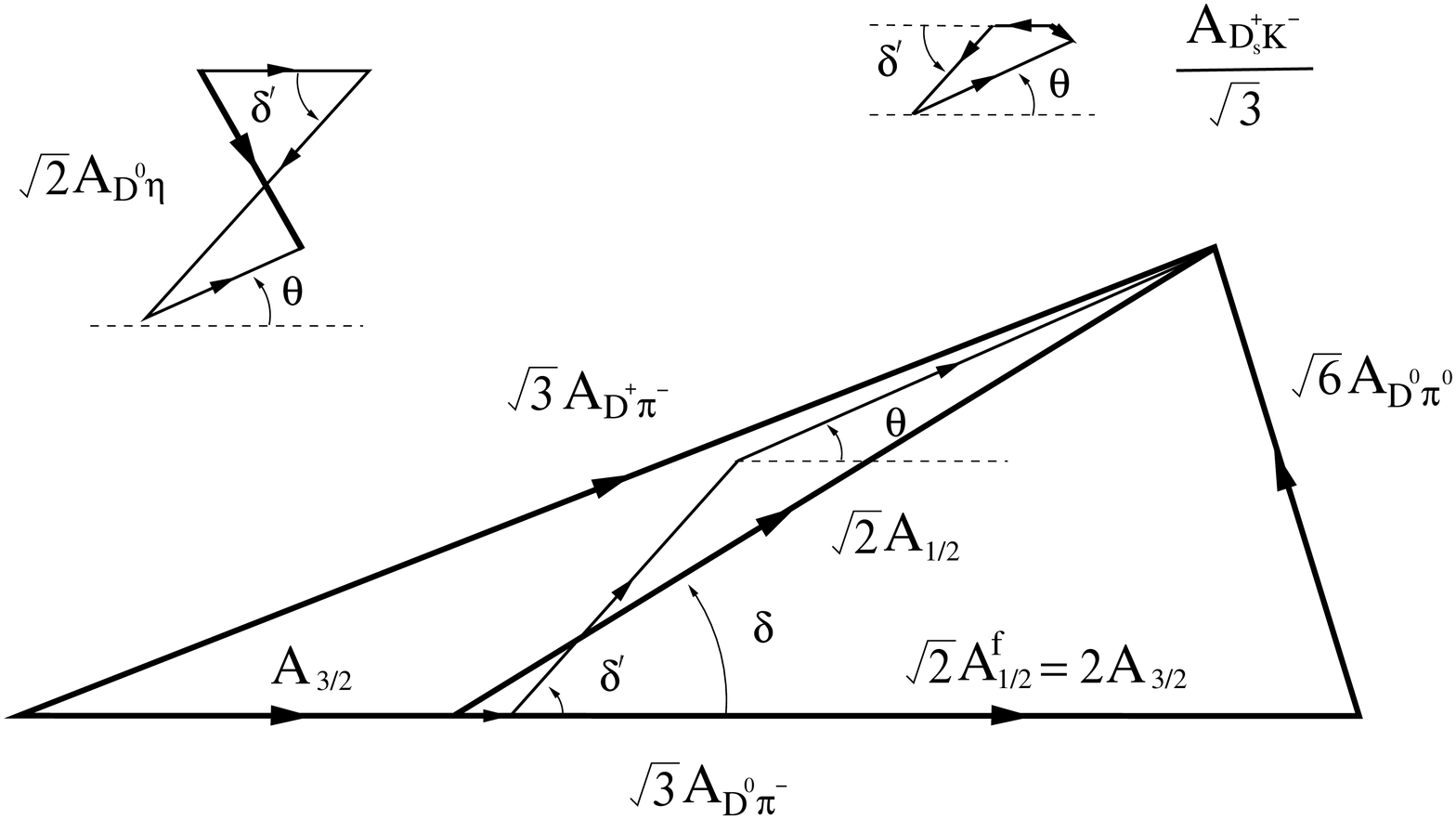 width 15cm)}
\smallskip
\caption{
Geometric representation of $\overline B{} \to DP$ rescattering
of Eq. (B1).
}
\label{fig:FSI1}
\end{figure}

We illustrate the amplitudes of Fit1$_{DP}$, i.e.
$\delta^\prime=47.8^\circ$ and $\theta=24.8^\circ$,
in Fig. \ref{fig:FSI1}. The $D^*P$ case is similar.
We have chosen the $x$-axis to coincide with $A_{3/2}$.
Since $\delta^\prime$ and $\theta$ are of the same sign,
when considering only $D^0\pi^0$, there is no need for a large angle, and
$\delta^\prime\sim \theta \sim \delta = \delta_{1/2} - \delta_{3/2}
 \sim 30^\circ$ would have been good enough.
However, because $\theta$ has same sign as $\delta^\prime$,
as we can see from Fig. \ref{fig:FSI1},
the third phaser contributing to $A_{D^0\eta}$, i.e.
$3 e^{i\theta}/16 (e^{i\delta_{3/2}}\sqrt2 A_{1/2}^f)$,
turns back and tends to reduce $\sqrt2 A_{D^0\eta}$.
We therefore need to start with a larger $\delta^\prime$ to compensate.
Thus, the relatively large phase $\delta^\prime\sim 50^\circ$
is driven by the strength of the measured ${\mathcal B}(D^0\eta)$
{\it and} ${\mathcal B}(D^0\pi^0)$.
Note that in this case the $\delta^\prime$ and $\theta$ phases
compensate strongly for each other and $A_{D_s^+K^-}$ is
small compared to $A_{D^0\eta}$ and $A_{D^0\pi^0}$.

For Fit2 of $DP$ case, $\delta^\prime(= 17.1^\circ)$ and 
$\theta(=-52.7^\circ)$ are opposite in sign.
This favors the generation of $D^0\eta$ since the
effects of $\delta^\prime$ and $\theta$ add to each other.
However, we would need a large $\theta$ phase to
overcome the effect of $\delta^\prime$ to generate $D^0\pi^0$ mode
i.e. to account for $|\delta| = |\delta_{1/2} - \delta_{3/2}| \sim 30^\circ$.
Otherwise, we will have too small a $A_{D^0\pi^0}$ amplitude.
Same as the previous case, 
a large phase of order $50^\circ$ is needed to account for the 
strength of both the $D^0\pi^0$ and $D^0\eta$ modes.
Inspecting the generation of $A_{D_s^+K^-}$, however,
we see that the effects of $\delta^\prime$ and $\theta$
now add to each other, and would generate too large a
${\mathcal B}(D_s^+K^-)$ that is already ruled out by data,
hence the case is not plotted.

We also refrain from plotting the case for $DV$ modes for the following reasons.
First of all, because of relatively weaker rescattering,
the $a_2$ effect is more prominent than in $D^{(*)}P$ case.
Second,
some discussion is already given in Sec. IV.B, where comparison is made
between FSI and $a_2$ contributions.
Decomposing into $A_{1/2}$ and $A_{3/2}$ amplitudes does not make the case clearer.

\end{widetext}

\end{document}